\documentclass[12pt]{article}

\usepackage{amsfonts}
\usepackage{amsmath}
\usepackage{amsthm}
\usepackage{cite}
\usepackage{epsfig}
\usepackage{latexsym}
\usepackage{paralist}
\usepackage{fancyhdr}
\usepackage{graphicx}
\numberwithin{equation}{section}
\usepackage[vcentermath]{youngtab}
\usepackage{young}
\usepackage{ytableau}
\usepackage{etex}
\usepackage{braket}
\usepackage{float}
\usepackage{extarrows}
\usepackage{autobreak}

\usepackage{pict2e}   
\usepackage{animate}  

\usepackage{multirow}
\usepackage{bigdelim}
\usepackage{fancybox}
\usepackage{cals}

\def\be{\begin{equation}}
\def\ee{\end{equation}}
\def\bea{\begin{eqnarray}}
\def\eea{\end{eqnarray}}

\renewcommand{\thefootnote}{\fnsymbol{footnote}}

\begin{document}

\hfuzz=100pt
\title{{\Large \bf{Coulomb branch in 3d $\mathcal{N}=2$ $SU(N)_k$ Chern-Simons gauge theories with chiral matter content}  }}
\date{}
\author{ Keita Nii$^a$\footnote{keita.nii@yukawa.kyoto-u.ac.jp}
}
\date{\today}

\maketitle

\thispagestyle{fancy}
\cfoot{}
\renewcommand{\headrulewidth}{0.0pt}

\vspace*{-1cm}
\begin{center}
$^{a}${{\it Center for Gravitational Physics}}
\\{{\it Yukawa Institute for Theoretical Physics, Kyoto University
}}
\\ {{\it Kitashirakawa Oiwake-cho, Sakyo-Ku, Kyoto, Japan}}  

\end{center}

\begin{abstract}
We elaborate on quantum moduli spaces in 3d $\mathcal{N}=2$ $SU(N)_k$ Chern-Simons gauge theories with $F$ fundamental and $\bar{F}$ anti-fundamental matter fields. The quantum flat direction on the Coulomb branch differs so much from the classical one and from the one of the vector-like theories. In many cases, the Coulomb branch is parametrized by the dressed monopoles. As is found from the computation of the superconformal index, these dressed operators at first sight appear to be dressed by massive elementary fields which don't seem to contribute to the low-energy physics. We argue that these dressed fields can be interpreted as a non-abelian monopole dressed (or not dressed) by massless matter fields.
Based on this analysis, we will report on the s-confinement phases with non-trivial monopole operators, which is consistent with the duality proposals \cite{Aharony:2014uya, Aharony:2013dha}. Along these studies, we find that the duality reported in \cite{Aharony:2014uya} must be modified when $k=\pm \frac{1}{2}(F-\bar{F})$ in order to have a correct duality map of the baryonic operators.
\end{abstract}

\renewcommand{\thefootnote}{\arabic{footnote}}
\setcounter{footnote}{0}

\newpage
\tableofcontents 

\newpage

\section{Introduction}
In three-dimensional spacetime, the $U(1)$ gauge dynamics is very interesting because the 3d electromagnetic interaction becomes strongly-coupled in the far-infrared limit and allows confinement in some cases. This situation is quite different from the 4d case where the $U(1)$ dynamics becomes weak in the infrared limit. In 3d supersymmetric gauge theories, the Coulomb branch in the moduli space of vacua non-perturbatively receives non-trivial quantum corrections and its quantum structure is very different from the classical one \cite{Affleck:1982as, Aharony:1997bx, deBoer:1997kr, Intriligator:2013lca, Aharony:2015pla, Csaki:2014cwa, Amariti:2015kha}.   
Recently, we have obtained a deeper understanding of the Coulomb branch in 3d $\mathcal{N}=2$ supersymmetric gauge theories \cite{Aharony:2013dha, Aharony:2013kma}. This development enables us to derive various 3d Seiberg-like dualities from known 4d dualities \cite{Seiberg:1994bz, Seiberg:1994pq}. In addition to the dimensional reduction of the dualities, the 3d Seiberg-like dualities have been independently developed as well \cite{Giveon:2008zn, Niarchos:2008jb, Niarchos:2009aa, Kapustin:2011vz, Kapustin:2011gh, Benini:2011mf, Willett:2011gp, Aharony:2014uya, Hwang:2012jh, Hwang:2015wna, Park:2013wta, Hwang:2011qt, Benini:2017dud,Dimofte:2017tpi, Bashmakov:2018ghn, Benvenuti:2018bav, Amariti:2018wht}.

In this paper, we study the quantum structure of the Coulomb moduli space in the 3d $\mathcal{N}=2$ $SU(N)_k$ Chern-Simons gauge theory with $F$ fundamental and $\bar{F}$ anti-fundamental chiral multiplets. Since the theory is ``chiral'' in a sense that the theory has unequal numbers of quarks and anti-quarks, its low-energy dynamics allows a very complicated Coulomb moduli space. By carefully studying the monopole operators which parametrize the Coulomb flat directions, we find s-confinement phases which are described by meson, baryon and dressed monopole operators. In this paper, we also examine the dualities proposed in \cite{Aharony:2014uya, Aharony:2013dha, Park:2013wta} and give a correct mapping of the baryon and monopole operators under the duality. In this analysis, we find that the duality reported in \cite{Aharony:2014uya} is incorrect for $k=\pm \frac{1}{2}(F-\bar{F})$ due to the mismatch of the anti-baryonic operator. We resolve this problem by proposing the correct duality for $k=\pm \frac{1}{2}(F-\bar{F})$. We will also discuss how this new duality appears from the duality known in the literature \cite{Aharony:2013dha}.

The rest of this paper is organized as follows: In Section 2, we will review the results of the 3d $\mathcal{N}=2$ $SU(2)_k$ Chern-Simons gauge theory with doublet quarks by focusing on the confinement phases, where there is no Coulomb branch. From Section 3 to Section 5, we will generalize the $SU(2)$ analysis to the $SU(3)$, $SU(4)$ and $SU(5)$ gauge groups, where the (dressed) monopole operator must be taken into account. In Section 6, we will study the confinement phases of the 3d $\mathcal{N}=2$ $SU(N)_k$ gauge theory. In Section 7, we study the baryonic and Coulomb branch directions of the moduli space and find a correct duality transformation of them. We will propose a modified duality for $k=\frac{1}{2}(F-\bar{F})$. In Section 8, we will summarize our findings.

\section{$SU(2)$ gauge group}
In this section, we will briefly review the results of the 3d $\mathcal{N}=2$ $SU(2)_k$ Chern-Simons (CS) gauge theories \cite{Intriligator:2013lca}. Since there is no difference between $SU(2)$ fundamental and anti-fundamental representations, we consider the $SU(2)_k$ gauge theory with $F$ doublets. The CS level $k$ is an integer for even $F$ whereas $k$ becomes a half-odd integer for odd $F$. Along the Coulomb branch, the gauge group is broken as $SU(2)_{k} \rightarrow U(1)_{2k}$ and the fundamental matter field is decomposed as $\mathbf{2} \rightarrow \mathbf{1}_1+ \mathbf{1}_{-1}$. Since the positively and negatively charged components are equally generated along this breaking, the CS term is not sifted at all. As a result, the low-energy dynamics of the $SU(2)$ Chern-Simons theory does not include the Coulomb branch which is massive. The Coulomb branch only appears for $SU(2)$ SQCD theories without a CS level. In what follows, we will list the s-confinement phases with a non-zero $k$, which was studied in \cite{Intriligator:2013lca}.

The first example is a 3d $\mathcal{N}=2$ $SU(2)_{k=\frac{1}{2}}$ gauge theory with three doublets, which is summarized in Table \ref{SU(2)3}. Since the total number of the doublets is odd, the Chern-Simons term must be a half-odd integer for maintaining the gauge-invariance. The bare Chern-Simons term is here tuned to be $k=\frac{1}{2}$. The Coulomb branch is massive since the bare CS level cannot be canceled along any RG flows. The theory is dual to a free meson $M_{i}:=Q^2$ which has a non-abelian flavor index.

\begin{table}[H]\caption{The 3d $\mathcal{N}=2$ $SU(2)_{k=\frac{1}{2}}$ gauge theory with three doublets} 
\begin{center}
  \begin{tabular}{|c||c|c|c|c| } \hline
  &$SU(2)_{k=1/2}$&$SU(3)$&$U(1)$&$U(1)_R$ \\ \hline
$Q$& ${\tiny \yng(1)}$&${\tiny \yng(1)}$&$1$&$r$ \\  \hline 
$M=Q^2$&1&${\tiny \overline{\yng(1)}}$&$2$&$2r$ \\ \hline
  \end{tabular}
  \end{center}\label{SU(2)3}
\end{table}

As a simple test of this duality, we can compare the superconformal indices \cite{Pestun:2007rz, Kapustin:2009kz, Hama:2010av}\cite{Bhattacharya:2008bja, Kim:2009wb, Imamura:2011su, Kapustin:2011jm} from the electric and s-confining descriptions. We find that the electric indices can be beautifully unified as follows:
\begin{align}
I_{SU(2)_{\frac{1}{2}}, 3 \, \Box } &=1+3 t^2 x^{1/2}+6 t^4 x+\left(10 t^6-\frac{3}{t^2}\right) x^{3/2}+\left(15 t^8-9\right) x^2+\left(21 t^{10}-15 t^2\right) x^{5/2} \nonumber \\
&+\left(28 t^{12}-21 t^4+\frac{3}{t^4}\right) x^3+\left(36 t^{14}-27 t^6+\frac{6}{t^2}\right) x^{7/2}+\left(45 t^{16}-33 t^8\right) x^4 \nonumber \\
&+\left(55 t^{18}-39 t^{10}-\frac{1}{t^6}-12 t^2\right) x^{9/2}+\left(66 t^{20}-45 t^{12}-24 t^4+\frac{6}{t^4}\right) x^5  \nonumber \\
&+\left(78 t^{22}-51 t^{14}-36 t^6+\frac{27}{t^2}\right) x^{11/2}+\left(91 t^{24}-57 t^{16}-48 t^8+44\right) x^6+\cdots \nonumber \\
&= \left. \left( \frac{(t^{-2} x^{2-2r};x^2)_\infty}{(t^2 x^{2r};x^2)_\infty} \right)^3 \right|_{r=\frac{1}{4}}, 
\end{align}
where the r-charge of $Q$ is set to be $r=\frac{1}{4}$ for simplicity and $t$ is a fugacity parameter for the global $U(1)$ symmetry. The second term $3 t^2 x^{1/2}$ corresponds to the meson operator $M:=QQ$ which has three components. On the last line, the index is represented as a single contribution of the free meson $M_i$, which confirms that the theory indeed exhibits s-confinement.

The next example is a 3d $\mathcal{N}=2$ $SU(2)_{k=1}$ gauge theory with two doublets, which is listed in Table \ref{SU(2)2}. Similarly, there is no Coulomb branch since the bare CS term cannot be canceled along the Coulomb branch. The low-energy physics is described by the meson $M:=Q^2$ which has one component. The dual description is a non-gauge theory with a free meson $M$.

\begin{table}[H]\caption{The 3d $\mathcal{N}=2$ $SU(2)_{k=1}$ gauge theory with two doublets} 
\begin{center}
  \begin{tabular}{|c||c|c|c|c| } \hline
  &$SU(2)_{k=1}$&$SU(2)$&$U(1)$&$U(1)_R$ \\ \hline
$Q$& ${\tiny \yng(1)}$&${\tiny \yng(1)}$&$1$&$r$ \\  \hline 
$M=Q^2$&1&1&$2$&$2r$ \\ \hline
  \end{tabular}
  \end{center}\label{SU(2)2}
\end{table}

As a check of this s-confinement phase, we will compute the superconformal indices from the electric and magnetic theories. The result is given by
\begin{align}
I_{SU(2)_{1}, 2 \, \Box }  &=1+t^2 x^{1/2}+t^4 x+\left(t^6-\frac{1}{t^2}\right) x^{3/2}+\left(t^8-1\right) x^2+t^{10} x^{5/2}+t^{12} x^3+\left(t^{14}-\frac{1}{t^2}\right) x^{7/2} \nonumber \\
&+\left(t^{16}-2\right) x^4+\left(t^{18}-t^2\right) x^{9/2}+\left(t^{20}+\frac{1}{t^4}\right) x^5+t^{22} x^{11/2}+\left(t^{24}-2\right) x^6 \nonumber \\
& +\left(t^{26}-2 t^2\right) x^{13/2}+\left(t^{28}-t^4+\frac{1}{t^4}\right) x^7+\left(t^{30}+\frac{1}{t^2}\right) x^{15/2}+\left(t^{32}-2\right) x^8+\cdots  \nonumber \\
&= \left. \frac{(t^{-2} x^{2-2r};x^2)_\infty}{(t^2 x^{2r};x^2)_\infty} \right|_{r=\frac{1}{4}},
\end{align}
where the parameter $t$ denotes fugacity for the $U(1)$ global symmetry. The r-charge is again set to be $r=\frac{1}{4}$ for simplicity. We can see that there is no contribution from the Coulomb branch operator. On the last line above, the index can be combined into a single contribution of the gauge-singlet chiral superfield $M$, which confirms the validity of our analysis.

\section{$SU(3)$ gauge group}
In this section, we consider the confinement phases in an $SU(3)$ Chern-Simons gauge theory with chiral (or vector-like) matter content. Although this paper mainly focuses on s-confinement phases with a monopole operator, this section also deals with s-confinement without a monopole for completeness. As in the $SU(2)$ case studied in \cite{Intriligator:2013lca}, we start with an s-confinement example of the SQCD theory without a CS level and deform it by real masses to flow into chiral theories. In many cases, the Coulomb branch will become massive due to those real masses but some cases allow a (dressed) monopole operator parametrizing the Coulomb moduli space.     

\subsection{$SU(3)_{k=0}$ with $(F,\bar{F})=(4,2)$}
The first example is a 3d $\mathcal{N}=2$ $SU(3)$ gauge theory with four fundamental and two anti-fundamental quarks (see Table \ref{SU(3)(4,2)0k}). The bare CS level is set to be zero. This theory was studied in \cite{Nii:2018bgf} and exhibits s-confinement. The Higgs branch is described by a meson $M:=Q \tilde{Q}$ and a baryon $B:=Q^3$. The bare Coulomb branch, denoted by $Y^{bare}$, induces a spontaneous breaking of the gauge symmetry
\begin{align}
SU(3) & \rightarrow U(1)_1 \times U(1)_2 \\
{\tiny \yng(1)}  & \rightarrow \mathbf{1}_{1,1}+\mathbf{1}_{0,-2}+\mathbf{1}_{-1,1} \\
{\tiny \overline{\yng(1)}} & \rightarrow \mathbf{1}_{-1,-1}+\mathbf{1}_{0,2}+\mathbf{1}_{1,-1}.
\end{align}
Along the Coulomb branch, the components charged under the $U(1)_1$ subgroup are massive and integrated out. As a result, the effective Chern-Simons terms are generated as follows:
\begin{align}
k_{eff}^{U(1)_1} =0,~~~~~~k_{eff}^{U(1)_1,U(1)_2} =F-\bar{F}=2.
\end{align}
Since the Coulomb branch $Y^{bare}$ corresponds to the $U(1)_1$ generator, the associated vector field cannot be massive. As can be seen above, the $U(1)_1$ CS term $k_{eff}^{U(1)_1}$ (which is a topological mass for the $U(1)_1$ vector field) correctly vanishes. However, due to the non-zero mixed Chern-Simons $k_{eff}^{U(1)_1,U(1)_2}$, the bare operator obtains a non-zero $U(1)_2$ charge \cite{Affleck:1982as, Intriligator:2013lca} and must be dressed by the matter fields \cite{Csaki:2014cwa, Amariti:2015kha, Aharony:2015pla ,Nii:2018bgf}
\begin{align}
Y_d:=Y^{bare}\mathbf{1}_{0,2} \sim Y^{bare} \tilde{Q}.
\end{align}
The quantum numbers of $Y_d$ can be computed as Table \ref{SU(3)(4,2)0k}.
The low-energy effective theory is described by $M,B$ and $Y_d$ with a cubic superpotential
\begin{align}
W_{eff}= BM Y_d.  \label{WeffSU(3)(4,2)0k}
\end{align}
In the rest of this section, we will derive various confinement phases by introducing real masses to some of $Q$ and $\tilde{Q}$. 

\begin{table}[H]\caption{The 3d $\mathcal{N}=2$ $SU(3)_{k=0}$ gauge theory with $(F,\bar{F})=(4,2)$} 
\begin{center}
  \begin{tabular}{|c||c|c|c|c|c|c| } \hline
  &$SU(3)_{k=0}$&$SU(4)$&$SU(2)$&$U(1)$&$U(1)$&$U(1)_R$ \\[3pt] \hline
$Q$& ${\tiny \yng(1)}$&$\tiny \yng(1)$&1&$1$&0&$0$ \\  
$\tilde{Q}$& ${\tiny \overline{\yng(1)}}$&1&$\tiny \yng(1)$&0&1&0  \\ \hline
$M:=Q\tilde{Q}$&1&${\tiny \yng(1)}$&$\tiny \yng(1)$&$1$&1&$0$ \\ 
$B:=Q^3$&1&${\tiny \overline{\yng(1)}}$&1&3&0&0  \\ \hline
$Y^{bare}$&$U(1)_1$: $0$, $U(1)_2:$ $-2$&1&1&$-4$&$-2$&$2$ \\
$Y_d:=Y^{bare} \tilde{Q}$&1&1&$\tiny \yng(1)$&$-4$&$-1$&$2$  \\ \hline
  \end{tabular}
  \end{center}\label{SU(3)(4,2)0k}
\end{table}

\subsection{$SU(3)_{k=\frac{1}{2}}$ with $(F,\bar{F})=(4,1)$}

We next study the 3d $\mathcal{N}=2$ $SU(3)$ gauge theory with four fundamental and one anti-fundamental matter fields. Since the total number of the (anti-)fundamental quarks is odd, the gauge group must have a half-odd integer CS level $k=\frac{1}{2}$. This theory can be obtained from the previous one by introducing a real mass to a single $\tilde{Q}^{\bar{f}=2}$. In the dual description, four components of the meson and a single component of $Y_d$ obtain real masses. Therefore, the present theory can have a Coulomb branch even in the presence of the CS term. This situation is quite different from the $SU(2)$ gauge theories. 

As in the previous case, we can define the dressed Coulomb branch operator based on $Y^{bare}$ although the explicit form of the resulting dressed operator seems to be composed of massive excitations. When the bare operator $Y^{bare}$ obtains a non-zero vev, the gauge group is spontaneously broken as 
\begin{align}
SU(3) & \rightarrow U(1)_1 \times U(1)_2 \\
{\tiny \yng(1)}  & \rightarrow \mathbf{1}_{1,1}+\mathbf{1}_{0,-2}+\mathbf{1}_{-1,1} \\
{\tiny \overline{\yng(1)}} & \rightarrow \mathbf{1}_{-1,-1}+\mathbf{1}_{0,2}+\mathbf{1}_{1,-1} \\
\mathbf{adj.} & \rightarrow \mathbf{1}_{0,0} +\mathbf{1}_{0,0} +\mathbf{1}_{1,3}+\mathbf{1}_{-1,-3}+\mathbf{1}_{2,0}+\mathbf{1}_{-2,0}+\mathbf{1}_{1,-3}+\mathbf{1}_{-1,3},      
\end{align}
where $\mathbf{adj.}$ corresponds to the gaugino field $W_{\alpha}$. In this case, the bare operator $Y^{bare}$ obtains both the $U(1)_1$ and $U(1)_2$ charges which are proportional to the effective CS terms $k_{eff}^{U(1)_1}$ and $k_{eff}^{U(1)_1,U(1)_2}$, respectively. See Table \ref{SU(3)(4,1)1/2k} for quantum numbers of $Y^{bare}$. In order to cancel these $U(1)$ charges of the bare operator $Y^{bare}$, we have to define
\begin{align}
Y_d:=Y^{bare} \mathbf{1}_{-1,3} \sim Y^{bare}  W_\alpha,
\end{align}
where $\mathbf{1}_{-1,3} $ is a component from the gaugino fields. Notice that under the monopole background, this gaugino component obtains an anomalous spin and is transmuted to a boson \cite{Polyakov:1988md, Dimofte:2011py, Aharony:2015pla}. Therefore, $Y_d$ is a bosonic operator and contributes to the superconformal index (SCI) with a plus sign. This construction of the dressed monopole is completely the same as the approach adopted in \cite{Aharony:2015pla} and consistent with the SCI interpretation. As claimed in \cite{Aharony:2015pla}, this dressed operator is written in terms of massive components which are charged under the $U(1)_1$ subgroup and gets masses along the Coulomb branch. In addition, the Coulomb branch itself is massive due to the effective CS term $k_{eff}^{U(1)_1}$. Therefore, from the low-energy viewpoint that is based on the moduli space, it is better to write down this operator in terms of the low-energy massless degrees of freedom. In what follows, we will show that this dressed operator can be represented by massless excitations along the quantum-mechanically flat moduli space.

When the Coulomb branch, denoted by $Y^{bare}_{SU(2)}$, obtains a non-zero expectation value, the gauge group is spontaneously broken to 
\begin{align}
SU(3) & \rightarrow SU(2) \times U(1) \\
{\tiny \yng(1)}  & \rightarrow {\tiny \yng(1)}_{\, 1} + \mathbf{1}_{-2}   \\
{\tiny \overline{\yng(1)}} & \rightarrow  {\tiny \yng(1)}_{\,-1} +\mathbf{1}_2,  
\end{align}
where the Coulomb branch operator $Y^{bare}_{SU(2)}$ is constructed from the $U(1)$ vector superfield by dualizing it. All the components are charged under the $U(1)_1$ subgroup and then massive along the Coulomb branch. 
Since there is a single $U(1)$ factor, there is no need to dress the bare monopole operator by matter multiplets. Since the matter content is ``chiral,'' various CS terms are generated along the Coulomb branch
\begin{align}
k_{eff}^{U(1)_1} &=6k -(F-\bar{F}) = 0\\
k^{SU(2)}_{eff} &= k+\frac{1}{2}(F-\bar{F}) = 2 
\end{align}
Since the CS term for $U(1)_1$ correctly vanishes, the Coulomb branch $Y_{SU(2)}^{bare}$ can be a flat direction. For the $SU(2)$ sector, its low-energy limit becomes a pure $SU(2)$ CS gauge theory with $k_{eff}^{SU(2)} =2$ whose Witten index allows a supersymmetric vacuum \cite{Witten:1982df, Witten:1999ds, Smilga:2012uy, Henningson:2012vb, Intriligator:2013lca}. In terms of the fundamental monopoles $Y_i \sim \exp(\phi_i -\phi_{i+1})$, the bare operator $Y^{bare}_{SU(2)}$ is defined by
\begin{align}
Y_{SU(2)}^{bare} :=  \sqrt{Y_1 Y_2^2},
\end{align}
which has the same quantum numbers as $Y_d$ and can be identified with it. The square root means that the monopole $Y_{SU(2)}^{bare}$ is a non-abelian monopole associated to the breaking $SU(3) \rightarrow SU(2) \times U(1)$ \cite{Preskill:1984gd, Weinberg:2012pjx, Csaki:2014cwa, Amariti:2015kha, Nii:2018bgf}. 
The low-energy dynamics is described by $M, B$ and $Y_d \sim Y_{SU(2)}^{bare}$ with an effective superpotential
\begin{align}
W_{eff}= BM Y_d.
\end{align}
We can easily derive this superpotential from the previous result \eqref{WeffSU(3)(4,2)0k} via a real mass deformation.

\begin{table}[H]\caption{The 3d $\mathcal{N}=2$ $SU(3)_{k=\frac{1}{2}}$ gauge theory with $(F,\bar{F})=(4,1)$} 
\begin{center}
  \begin{tabular}{|c||c|c|c|c|c| } \hline
  &$SU(3)_{k=\frac{1}{2}}$&$SU(4)$&$U(1)$&$U(1)$&$U(1)_R$ \\[3pt] \hline
$Q$& ${\tiny \yng(1)}$&$\tiny \yng(1)$&$1$&0&$0$ \\  
$\tilde{Q}$& ${\tiny \overline{\yng(1)}}$&1&0&1&0  \\ \hline
$M:=Q\tilde{Q}$&1&${\tiny \yng(1)}$&$1$&1&$0$ \\ 
$B:=Q^3$&1&${\tiny \overline{\yng(1)}}$&3&0&0  \\ \hline
$Y^{bare}$&$U(1)_1$: $1$, $U(1)_2:$ $-3$&1&$-4$&$-1$&$1$ \\
$Y_d:=Y^{bare} W_\alpha$&1&1&$-4$&$-1$&$2$  \\ \hline
  \end{tabular}
  \end{center}\label{SU(3)(4,1)1/2k}
\end{table}

Finally, we will confirm the agreement of the superconformal indices between the electric and magnetic descriptions. On both sides of the duality, the indices are computed as 
\begin{align}
I &=1+\frac{x^{1/3}}{t^4 u}+x^{2/3} \left(\frac{1}{t^8 u^2}+4 t u\right)+x \left(\frac{1}{t^{12} u^3}+4 t^3\right)+x^{4/3} \left(\frac{1}{t^{16} u^4}+10 t^2 u^2\right) \nonumber \\
& +x^{5/3} \left(\frac{1}{t^{20} u^5}+15 t^4 u-\frac{6 u}{t^2}\right)+x^2 \left(\frac{1}{t^{24} u^6}+10 t^6+20 t^3 u^3-17\right) \nonumber \\
& +x^{7/3} \left(\frac{1}{t^{28} u^7}+36 t^5 u^2-\frac{6 t^2}{u}-\frac{20 u^2}{t}\right)+x^{8/3} \left(\frac{1}{t^{32} u^8}+36 t^7 u+\frac{4 u}{t^5}+35 t^4 u^4-60 t u\right) \nonumber \\
& +x^3 \left(\frac{1}{t^{36} u^9}+20 t^9+70 t^6 u^3-60 t^3+\frac{28}{t^3}-45 u^3\right)+\cdots \nonumber \\
& = \left.  \left( \frac{(t^{-1}u^{-1} x^{2-2r};x^2)_\infty}{(tu x^{2r};x^2)_\infty} \right)^4  \left(  \frac{(t^{-3} x^{2-3r};x^2)_\infty}{(t^3 x^{3r};x^2)_\infty} \right)^4 \left( \frac{(t^{4}u^{1} x^{5r};x^2)_\infty}{(t^{-4}u^{-1} x^{2-5r};x^2)_\infty}  \right)^1  \right|_{r=\frac{1}{3}},
\end{align} 
where the r-charges of the electric elementary fields are set to be $r=r_Q=r_{\tilde{Q}}=\frac{1}{3}$ for simplicity. The fugacity parameters $t$ and $u$ account for the two $U(1)$ global symmetries. In the last line above, we have rewritten the indices in terms of the three chiral multiplets $M,B$ and $Y_d$, which confirms the validity of our analysis.

\subsection{$SU(3)_{k=1}$ with $(F,\bar{F})=(4,0)$}
For completeness of this section, we also study the confinement phases without a Coulomb branch operator. The UV description is a 3d $\mathcal{N}=2$ $SU(3)_{k=1}$ gauge theory with four fundamental chiral multiplets and a level-1 CS term (see Table \ref{SU(3)(4,0)1k}). Since the theory is completely chiral, the Higgs branch is described by the baryon operator $B:=Q^3$. In any breaking patterns of the gauge group along the Coulomb moduli space, which is realized by tuning the vev of the adjoint scalar in the vector multiplet, the corresponding Coulomb branch obtains a non-zero CS term. As a result, there is no Coulomb moduli space in this example. The low-energy description is equivalent to a free baryon $B:=Q^3$ which has four components. This result is consistent with the confinement phase in Table \ref{SU(3)(4,1)1/2k} via a real mass deformation to $\tilde{Q}$. 

\begin{table}[H]\caption{The 3d $\mathcal{N}=2$ $SU(3)_{k=1}$ gauge theory with $(F,\bar{F})=(4,0)$} 
\begin{center}
  \begin{tabular}{|c||c|c|c|c| } \hline
  &$SU(3)_{k=1}$&$SU(4)$&$U(1)$&$U(1)_R$ \\[3pt] \hline
$Q$& ${\tiny \yng(1)}$&$\tiny \yng(1)$&$1$&$0$ \\   \hline
$B:=Q^3$&1&${\tiny \overline{\yng(1)}}$&3&0  \\ \hline
  \end{tabular}
  \end{center}\label{SU(3)(4,0)1k}
\end{table}

As a consistency check, we will compute the superconformal indices. We define $t$ as a fugacity parameter for the global $U(1)$ symmetry and set the r-charge of $Q$ to be $r =\frac{1}{3}$. The indices are expanded as
\begin{align}
I_{SU(3)_{k=1},(4,0)} &=1+\left( 4t^3 -\frac{4}{t^3} \right)x+\left(10 t^6+\frac{6}{t^6}-16\right) x^2+\left(20 t^9-\frac{4}{t^9}-36 t^3+\frac{20}{t^3}\right) x^3 \nonumber \\
&+\left(35 t^{12}+\frac{1}{t^{12}}-64 t^6+28\right) x^4+\left(56 t^{15}-100 t^9-\frac{20}{t^9}+20 t^3+\frac{44}{t^3}\right) x^5+\cdots \nonumber \\
&= \left. \left( \frac{(t^{-3} x^{2-3r};x^2)_\infty}{(t^3 x^{3r};x^2)_\infty}  \right)^{4}   \right|_{r=\frac{1}{3}},
\end{align}
where in the last line, the index is rewritten in terms of the indices of the four free chiral superfields.
Notice that the 3d $\mathcal{N}=2$ $SU(3)_{k=0}$ gauge theory with $(F,\bar{F})=(4,0)$ also leads to the free baryon $B:=Q^3$ \cite{Aharony:2013dha, Nii:2018bgf} and gives the same superconformal indices although each contribution in the SCI comes from the state with different GNO charges.

\subsection{$SU(3)_{k=\frac{1}{2}}$ with $(F,\bar{F})=(3,2)$}
The electric description is a 3d $\mathcal{N}=2$ $SU(3)$ gauge theory with $(F,\bar{F})=(3,2)$ (anti-)fundamental matter fields, which is obtained from the SQCD theory in Table \ref{SU(3)(4,2)0k} via a real mass to a fundamental quark $Q^{F=4}$. On the dual side, the corresponding real mass makes all the components of the dressed monopole $Y_d$ massive. Therefore, there is no Coulomb branch in this theory. The absence of the Coulomb branch can be more directly seen as follows: Along the Coulomb branch $Y_{SU(2)}$, which induces the gauge symmetry breaking $SU(3) \rightarrow SU(2) \times U(1)$, the effective CS level becomes $k_{eff}^{U(1)} =6k -(F-\bar{F}) \neq 0$ which corresponds to a topological mass term for the vector multiplet. For other Coulomb branches, we can give a similar argument.
Therefore, there is no Coulomb branch and the low-energy dynamics is described by $M:=Q \tilde{Q}$ and $B:=Q^3$. The quantum numbers of these fields are summarized in Table \ref{SU(3)(3,2)1/2k}. Since we cannot write down any superpotential from $M$ and $B$, we conclude that $M$ and $B$ become free fields in the far-infrared limit. 

\begin{table}[H]\caption{The 3d $\mathcal{N}=2$ $SU(3)_{k=\frac{1}{2}}$ gauge theory with $(F,\bar{F})=(3,2)$} 
\begin{center}
  \begin{tabular}{|c||c|c|c|c|c|c| } \hline
  &$SU(3)_{k=\frac{1}{2}}$&$SU(3)$&$SU(2)$&$U(1)$&$U(1)$&$U(1)_R$ \\[3pt] \hline
$Q$& ${\tiny \yng(1)}$&$\tiny \yng(1)$&1&$1$&0&$0$ \\  
$\tilde{Q}$& ${\tiny \overline{\yng(1)}}$&1&$\tiny \yng(1)$&0&1&0  \\ \hline
$M:=Q\tilde{Q}$&1&${\tiny \yng(1)}$&$\tiny \yng(1)$&$1$&1&$0$ \\ 
$B:=Q^3$&1&$1$&1&3&0&0  \\ \hline
  \end{tabular}
  \end{center}\label{SU(3)(3,2)1/2k}
\end{table}

As a test of the duality, we compute the superconformal indices. The result is given by 
\begin{align}
I_{SU(3)_{k=\frac{1}{2}},(3,2)} &=1+6 t u x^{2/3}+\left(t^3-\frac{1}{t^3}\right) x+x^{4/3} \left(21 t^2 u^2-\frac{6}{t u}\right)+x^{5/3} \left(6 t^4 u-\frac{6 u}{t^2}\right) \nonumber \\
&+x^2 \left(t^6+56 t^3 u^3-37\right)+\frac{3 x^{7/3} \left(7 t^9 u^3-2 t^6-7 t^3 u^3+2\right)}{t^4 u}\nonumber \\
&+x^{8/3} \left(6 t^7 u+126 t^4 u^4+\frac{15}{t^2 u^2}-126 t u\right) \nonumber \\
&+x^3 \left(t^9+56 t^6 u^3-36 t^3+\frac{35}{t^3}-56 u^3\right)+\cdots  \nonumber \\
&=  \left.  \left( \frac{(t^{-1}u^{-1} x^{2-2r};x^2)_\infty}{(tu x^{2r};x^2)_\infty} \right)^6  \left(  \frac{(t^{-3} x^{2-3r};x^2)_\infty}{(t^3 x^{3r};x^2)_\infty} \right)^1 \right|_{r=\frac{1}{3}},
\end{align}
where we set the r-charges to be $r=r_{Q}=r_{\tilde{Q}} =\frac{1}{3}$ for simplicity. $t$ and $u$ are the fugacity parameters for the two $U(1)$ global symmetries. The second term $6 t u x^{2/3}$ is identified with the meson $M:=Q \tilde{Q}$. The baryon $B:=Q^3$ appears as $t^3 x$ in the index. The higher-order terms are the symmetric products of these bosons and the fermion contributions. In the last line, the index can be unified into two contributions of the meson and baryon chiral multiplets, which confirms the validity of our analysis.

\subsection{$SU(3)_{k=1}$ with $(F,\bar{F})=(3,1)$}
By adding a real mass to an anti-fundamental matter field in the previous theory (Table \ref{SU(3)(3,2)1/2k}), we can flow to the 3d $\mathcal{N}=2$ $SU(3)$ gauge theory with $(F,\bar{F})=(3,1)$ (anti-)fundamental chiral multiplets. On the dual side, the corresponding real mass lifts some of the meson components. The resulting low-energy degrees of freedom are described by $M:=Q \tilde{Q}$ and $B:=Q^3$. This theory also shows s-confinement without a monopole operator. 
The quantum numbers of the moduli fields are summarized in Table \ref{SU(3)(3,1)1k}.

\begin{table}[H]\caption{The 3d $\mathcal{N}=2$ $SU(3)_{k=1}$ gauge theory with $(F,\bar{F})=(3,1)$} 
\begin{center}
  \begin{tabular}{|c||c|c|c|c|c| } \hline
  &$SU(3)_{k=1}$&$SU(3)$&$U(1)$&$U(1)$&$U(1)_R$ \\[3pt] \hline
$Q$& ${\tiny \yng(1)}$&$\tiny \yng(1)$&$1$&0&$0$ \\  
$\tilde{Q}$& ${\tiny \overline{\yng(1)}}$&1&0&1&0  \\ \hline
$M:=Q\tilde{Q}$&1&${\tiny \yng(1)}$&$1$&1&$0$ \\   
$B:=Q^3$&1&1&3&0&0  \\ \hline
  \end{tabular}
  \end{center}\label{SU(3)(3,1)1k}
\end{table}

It is straightforward to check that the superconformal indices can be combined into the contributions of the two chiral superfields $M$ and $B$:
\begin{align}
I_{SU(3)_{k=1},(3,1)}&=1+3 t u x^{2/3}+\left(t^3-\frac{1}{t^3}\right) x+x^{4/3} \left(6 t^2 u^2-\frac{3}{t u}\right)+x^{5/3} \left(3 t^4 u-\frac{3 u}{t^2}\right) \nonumber \\
&+x^2 \left(t^6+10 t^3 u^3-10\right)+\frac{3 x^{7/3} \left(2 t^9 u^3-t^6-2 t^3 u^3+1\right)}{t^4 u}\nonumber \\
&+x^{8/3} \left(3 t^7 u+15 t^4 u^4+\frac{3}{t^2 u^2}-18 t u\right)+x^3 \left(t^9+10 t^6 u^3-9 t^3+\frac{8}{t^3}-10 u^3\right)+\cdots \nonumber \\
&= \left.  \left( \frac{(t^{-1}u^{-1} x^{2-2r};x^2)_\infty}{(tu x^{2r};x^2)_\infty} \right)^3  \left(  \frac{(t^{-3} x^{2-3r};x^2)_\infty}{(t^3 x^{3r};x^2)_\infty} \right)^1 \right|_{r=\frac{1}{3}},
\end{align}
where $r$-charges are set to be $r_Q=r_{\tilde{Q}}=\frac{1}{3}$ for simplicity. The fugacity parameters $t$ and $u$ account for the two $U(1)$ global symmetries. The second term $3 t u x^{2/3}$ corresponds to the meson $M:=Q \tilde{Q}$ and the third term $t^3 x$ is identified with the baryon $B:=Q^3$. The other contributions are symmetric product of these bosons and the fermion contributions. 
In the last line, the indices are written in terms of $M$ and $B$, which confirms our analysis.

\subsection{$SU(3)_{k=\frac{3}{2}}$ with $(F,\bar{F})=(3,0)$}
By adding a positive real mass to an anti-fundamental quark in the previous theory, we obtain a 3d $\mathcal{N}=2$ $SU(3)_{k=\frac{3}{2}}$ gauge theory with three fundamental chiral multiplets. In the dual description, the corresponding real mass makes the meson massive and just removes it. There is no Coulomb branch operator in this case as well. The low-energy dynamics is described by a free baryon $B:=Q^3$. The quantum numbers of the moduli and elementary fields are summarized in Table \ref{SU(3)(3,0)3/2}.

We should notice that the low-energy dynamics drastically changes for a different value of $k$. As an example, we consider an $SU(3)_{k=\frac{1}{2}}$ gauge theory with $(F,\bar{F})=(3,0)$. In this case, there is a Coulomb branch $Y_{SU(2)}$ which corresponds to the abelian generator $t_{U(1)}=\mathrm{diag.}(1,1,-2)$ and its vev breaks the gauge group as $SU(3) \rightarrow SU(2) \times U(1)$. In this case, the effective CS terms become
\begin{align}
k_{eff}^{U(1)}=6k-(F-\bar{F})=0,~~~~~k_{eff}^{SU(2)}=k+\frac{1}{2}(F-\bar{F})=2,
\end{align}
which means that the $U(1)$ flat direction is indeed massless and that the low-energy $SU(2)$ theory has a supersymmetric vacuum \cite{Intriligator:2013lca} (non-zero Witten index). The low-energy dynamics is described by a single constraint $BY_{SU(2)}=1$. This constraint can be derived from the $SU(3)_{k=\frac{1}{2}}$ gauge theory with $(F,\bar{F})=(4,1)$ via a complex mass deformation to a single flavor. We can see that the present case with $k=\frac{3}{2}$ leads to a non-zero $k_{eff}^{U(1)}$ and this flat direction is not available.

\begin{table}[H]\caption{The 3d $\mathcal{N}=2$ $SU(3)_{k=\frac{3}{2}}$ gauge theory with $(F,\bar{F})=(3,0)$} 
\begin{center}
  \begin{tabular}{|c||c|c|c|c| } \hline
  &$SU(3)_{k=\frac{3}{2}}$&$SU(3)$&$U(1)$&$U(1)_R$ \\[3pt] \hline
$Q$& ${\tiny \yng(1)}$&$\tiny \yng(1)$&$1$&$0$ \\   \hline
$B:=Q^3$&1&$1$&3&0  \\ \hline
  \end{tabular}
  \end{center}\label{SU(3)(3,0)3/2}
\end{table}

Finally, we will test the superconformal indices of the electric and confining descriptions. It is straightforward to show that the superconformal index becomes
\begin{align}
I_{SU(3)_{k=\frac{3}{2}}, (3,0)} &=1+\left(t^3-\frac{1}{t^3}\right) x+\left(t^6-1\right) x^2+\left(t^9-\frac{1}{t^3}\right) x^3\nonumber \\
& \qquad +\left(t^{12}+\frac{1}{t^6}-2\right) x^4+\left(t^{15}-t^3\right) x^5+\cdots \nonumber \\
&=  \left. \left( \frac{(t^{-3} x^{2-3r};x^2)_\infty}{(t^3 x^{3r};x^2)_\infty}  \right)^{1}   \right|_{r=\frac{1}{3}},
\end{align}
where we set the r-charge of $Q$ to be $r=\frac{1}{3}$ and $t$ is a fugacity parameter for the global $U(1)$ symmetry. The baryon $B:=Q^3$ is represented as $t^3 x$. The indices can be unified into a single contribution of the gauge-singlet chiral superfield $B$. This supports our analysis.

\subsection{$SU(3)_{k=1}$ with $(F,\bar{F})=(2,2)$}
Let us move on to the vector-like example: The ultraviolet description is a 3d $\mathcal{N}=2$ $SU(3)_{k=1}$ gauge theory with $(F,\bar{F})=(2,2)$ (anti-)fundamental chiral multiplets. This theory can be obtained from an $SU(3)_{k=\frac{1}{2}}$ with $(F,\bar{F})=(3,2)$ via a real mass deformation to a single $Q_{f=3}$. Since the parent theory has no Coulomb branch operator, this daughter theory has no Coulomb branch as well. As a result, the low-energy theory is described by a meson $M:=Q \tilde{Q}$ whose quantum numbers are summarized in Table \ref{SU(3)(2,2)1k}. 

The absence of the Coulomb branch can be more concretely understood as follows: Along the Coulomb branch $Y_{SU(2)}^{bare}$, which induces the gauge symmetry breaking $SU(3) \rightarrow SU(2) \times U(1)$, the effective CS terms are computed as
\begin{align}
k_{eff}^{U(1)} = 6k=6,~~~k_{eff}^{SU(2)}=k=1.
\end{align}
Notice that the CS level shifts generated by the one-loop diagrams of the massive fermions are canceled for vector-like theories. 
Since the $U(1)$ CS term $k_{eff}^{U(1)}$ does not vanish, the Coulomb branch is lifted (massive). In addition, the low-energy pure $SU(2)_{k_{eff}^{SU(2)}=1}$ CS theory has no supersymmetric vacuum \cite{Intriligator:2013lca}. Therefore, this Coulomb branch $Y_{SU(2)}^{bare}$ is eliminated from the quantum moduli space. We can repeat the same argument for other directions of the Coulomb branch. As a result, there is no Coulomb moduli space in this theory.

\begin{table}[H]\caption{The 3d $\mathcal{N}=2$ $SU(3)_{k=1}$ gauge theory with $(F,\bar{F})=(2,2)$} 
\begin{center}
  \begin{tabular}{|c||c|c|c|c|c|c| } \hline
  &$SU(3)_{k=1}$&$SU(2)$&$SU(2)$&$U(1)$&$U(1)$&$U(1)_R$ \\[3pt] \hline
$Q$& ${\tiny \yng(1)}$&$\tiny \yng(1)$&1&$1$&0&$0$ \\  
$\tilde{Q}$& ${\tiny \overline{\yng(1)}}$&1&$\tiny \yng(1)$&0&1&0  \\ \hline
$M:=Q\tilde{Q}$&1&${\tiny \yng(1)}$&$\tiny \yng(1)$&$1$&1&$0$ \\  \hline
  \end{tabular}
  \end{center}\label{SU(3)(2,2)1k}
\end{table}

As a consistency check of our analysis, we will compute the superconformal indices. The result is given by
\begin{align}
I_{SU(3)_{k=1},(2,2)} &=1+4 t u x^{2/3}+x^{4/3} \left(10 t^2 u^2-\frac{4}{t u}\right)  \nonumber \\
&\qquad +x^2 \left(20 t^3 u^3-16\right)+x^{8/3} \left(35 t^4 u^4+\frac{6}{t^2 u^2}-36 t u\right)+ \cdots  \nonumber \\
&=\left. \left( \frac{(t^{-1} u^{-1} x^{2-2r};x^2)_\infty}{(t u x^{2r};x^2)_\infty}  \right)^{4}   \right|_{r=\frac{1}{3}},
\end{align}
where the r-charges are set to be $r=r_{Q}=r_{\bar{Q}} =\frac{1}{3}$ for simplicity. We can see that the index only contains the meson contributions and it is possible to sum up the perturbative series into the index of the four chiral superfields as in the last line.

\subsection{$SU(3)_{k=\frac{3}{2}}$ with $(F,\bar{F})=(2,1)$}
The next example is a 3d $\mathcal{N}=2$ $SU(3)$ gauge theory with $(F,\bar{F})=(2,1)$ (anti-)fundamental matter fields (see Table \ref{SU(3),(2,2),3/2k}), which can be obtained from the previous example via a real mass deformation for $\tilde{Q}^{\bar{f}=2}$. Therefore, there is no Coulomb branch in this case as well. This is consistent with the fact that the effective CS term associated with the breaking $SU(3) \rightarrow SU(2) \times U(1)$, whose $U(1)$ generator is $t_{U(1)}=\mathrm{diag.}(1,1,-2)$, cannot be non-zero:
\begin{align}
k_{eff}^{U(1)} =6k -(F -\bar{F}) =8.
\end{align}
This flat direction obtains a topological mass term proportional to $k_{eff}^{U(1)}$.
Similarly, for other directions of the Coulomb branch, we cannot find any $U(1)$ direction with vanishing $k_{eff}^{U(1)}$. This explains the absence of the Coulomb moduli space.  
The low-energy theory is then described by a single meson $M:=Q \tilde{Q}$.

\begin{table}[H]\caption{The 3d $\mathcal{N}=2$ $SU(3)_{k=\frac{3}{2}}$ gauge theory with $(F,\bar{F})=(2,1)$} 
\begin{center}
  \begin{tabular}{|c||c|c|c|c|c| } \hline
  &$SU(3)_{k=\frac{3}{2}}$&$SU(2)$&$U(1)$&$U(1)$&$U(1)_R$ \\[3pt] \hline
$Q$& ${\tiny \yng(1)}$&$\tiny \yng(1)$&$1$&0&$0$ \\  
$\tilde{Q}$& ${\tiny \overline{\yng(1)}}$&1&0&1&0  \\ \hline
$M:=Q\tilde{Q}$&1&${\tiny \yng(1)}$&$1$&1&$0$ \\  \hline 
  \end{tabular}
  \end{center}\label{SU(3),(2,2),3/2k}
\end{table}

Let us compare the superconformal indices calculated from the electric and confining descriptions. Both of the descriptions give the same index as
\begin{align} 
I_{SU(3)_{k=\frac{3}{2}}, (2,1)} &=1+2 t u x^{2/3}+x^{4/3} \left(3 t^2 u^2-\frac{2}{t u}\right)+x^2 \left(4 t^3 u^3-4\right) \nonumber \\
&+x^{8/3} \left(5 t^4 u^4+\frac{1}{t^2 u^2}-4 t u\right)+x^{10/3} \left(6 t^5 u^5-4 t^2 u^2\right)+x^4 \left(7 t^6 u^6-4 t^3 u^3-5\right)+\cdots \nonumber \\
&=  \left. \left( \frac{(t^{-1} u^{-1} x^{2-2r};x^2)_\infty}{(t u x^{2r};x^2)_\infty}  \right)^{2}   \right|_{r=\frac{1}{3}},
\end{align}
where $t$ and $u$ are the fugacity parameters for the global $U(1)$ symmetries. The r-charges of $Q$ and $\tilde{Q}$ are set to be $r=\frac{1}{3}$ for simplicity. This index can be recast into the index of a free meson which has two components. This confirms that the low-energy description is equivalent to a free meson $M:=Q \tilde{Q}$.

\subsection{$SU(3)_{k=2}$ with $(F,\bar{F})=(1,1)$}
The final example is a 3d $\mathcal{N}=2$ $SU(3)_{k=2}$ gauge theory with a single flavor (Table \ref{SU(3)11,2k}). The bare Chern-Simons term has to be integer for vector-like theories and we here take $k=2$. The Higgs branch is now simple enough and described by a single meson $M:=Q \tilde{Q}$. The low-energy dynamics of this theory can be obtained from the result of the previous subsection via a real mass deformation to a single fundamental quark $Q^{f=2}$. Hence, there is again no Coulomb branch. 
Roughly speaking, the Chern-Simons term gives masses to all the Coulomb directions and removes them from the chiral ring elements. We should notice that the magnitude of the Chern-Simons level is important for the low-energy physics: Since the gauge group is broken to $SU(2)$ along the Higgs branch $\braket{M}  \neq 0$, the low-energy pure $SU(2)$ gauge theory must have a supersymmetric vacuum. This is possible for $k \ge 2$ \cite{Intriligator:2013lca}. The low-energy effective theory for $k=2$ is dual to a free meson. Furthermore, we can claim that the $SU(3)_{k < 2}$ with $(F,\bar{F})=(1,1)$ spontaneously breaks supersymmetry.

\begin{table}[H]\caption{The 3d $\mathcal{N}=2$ $SU(3)_{k=2}$ gauge theory with $(F,\bar{F})=(1,1)$} 
\begin{center}
  \begin{tabular}{|c||c|c|c|c| } \hline
  &$SU(3)_{k=2}$&$U(1)$&$U(1)$&$U(1)_R$ \\[3pt] \hline
$Q$& ${\tiny \yng(1)}$&1&$0$&$0$ \\   
$\tilde{Q}$&${\tiny \overline{\yng(1)}}$&0&1&0  \\ \hline
$M:=Q \tilde{Q}$&1&1&1&0  \\ \hline
  \end{tabular}
  \end{center}\label{SU(3)11,2k}
\end{table}

As a test of our analysis, we compute the superconformal indices of the electric and s-confining descriptions. We observed a nice agreement of the indices as follows:
\begin{align}
I_{SU(3)_{k=2},(1,1)} &=1+t u x^{2/3}+x^{4/3} \left(t^2 u^2-\frac{1}{t u}\right)+x^2 \left(t^3 u^3-1\right) \nonumber \\
&\qquad +t^4 u^4 x^{8/3}+x^{10/3} \left(t^5 u^5-\frac{1}{t u}\right) +x^4 \left(t^6 u^6-2\right)+ \cdots \nonumber \\
& =  \left. \left( \frac{(t^{-1} u^{-1} x^{2-2r};x^2)_\infty}{(t u x^{2r};x^2)_\infty}  \right)^{1}   \right|_{r=\frac{1}{3}},
\end{align}
where $t$ and $u$ are the fugacity parameters of the two $U(1)$ global symmetries. The r-charges of $Q$ and $\tilde{Q}$ are set to be $r=r_{Q}=r_{\bar{Q}} =\frac{1}{3}$ for simplicity. In the first line, the indices are perturbatively calculated and these are summed into a single contribution of the gauge-singlet chiral multiplet $M$, which confirms that the theory in Table \ref{SU(3)11,2k} shows s-confinement. For the $SU(3)_{k=1}$ gauge theory with $(F,\bar{F})=(1,1)$, we observed that the SCI vanishes, which implies the supersymmetry breaking. This is consistent with our analysis.

\section{$SU(4)$ gauge group}
In this section, we study the 3d $\mathcal{N}=2$ $SU(4)_k$ Chern-Simons gauge theory with chiral matter content. We here focus on the s-confinement phases with (dressed) monopoles although there are also phases of the quantum deformed moduli space with monopole operators. Both of these phases will be generalized to an $SU(N)$ gauge symmetry in Section 6. 

\subsection{$SU(4)_{k=\frac{1}{2}}$ with $(F, \bar{F})=(5,2)$}
As studied in \cite{Nii:2018bgf}, it is known that the 3d $\mathcal{N}=2$ $SU(4)_{k=0}$ with $(F, \bar{F})=(5,3)$ (anti-) fundamental quarks exhibits s-confinement. Therefore, the 3d $\mathcal{N}=2$ $SU(4)_{k=\frac{1}{2}}$ with $(F, \bar{F})=(5,2)$, which is obtained by a real mass deformation to an anti-fundamental multiplet from the above theory, also shows s-confinement. On the magnetic (s-confining) side, the corresponding real mass keeps the (dressed) Coulomb branch massless and then we expect that this theory has a non-trivial monopole even in the presence of the non-zero CS term. As we will see below, we need to consider an asymmetric breaking pattern on the Coulomb branch.
The Higgs branch is described by
\begin{align}
M:=Q \tilde{Q},~~~~~~B:=Q^4.
\end{align}
The elementary and moduli fields are summarized in Table \ref{SU4,52,hk}. The classical dimension of the Higgs branch is $13$ and the total number of the above operators is $15$. As we will see below, this difference is correctly reduced by an $F$-flatness condition of the effective superpotential.

We claim that the following Coulomb branch can survive all the quantum corrections: The Coulomb branch $Y_{SU(2)}^{bare}$ is realized by turning on a vev of the adjoint scalar $\braket{\phi_{adj.}} =\mathrm{diag.} (v,v,0,-2v) $. Along this flat direction, the gauge group is broken as
\begin{align}
SU(4) & \rightarrow SU(2) \times U(1)_1 \times U(1)_2 \\ 
t_{U(1)_1}&:= \mathrm{diag.} (1,1,0,-2),~~~t_{U(1)_2}:= \mathrm{diag.} (0,0,1,-1)  \\
{\tiny \yng(1)} & \rightarrow  {\tiny \yng(1)}_{ \,1,0} + \mathbf{1}_{0,1} +\mathbf{1}_{-2,-1} \\
{\tiny \overline{\yng(1)}} & \rightarrow  {\tiny \yng(1)}_{ \,-1,0} +\mathbf{1}_{0,-1} +\mathbf{1}_{2,1} \\
\mathbf{adj.}   & \rightarrow  \mathbf{adj.}_{0,0}+\mathbf{1}_{0,0}+\mathbf{1}_{0,0}+\mathbf{1}_{2,2}+\mathbf{1}_{-2,-2} +  {\tiny \yng(1)}_{ \,1,-1}+ {\tiny \yng(1)}_{ \,3,1}+ {\tiny \yng(1)}_{ \,-1,1}+ {\tiny \yng(1)}_{ \,-3,-1}  ,
\end{align}
where the Coulomb branch $Y_{SU(2)}^{bare}$ is associated with the abelian generator $t_{U(1)_1}$ and the components charged under the $U(1)_1$ symmetry are massive. By integrating out the massive fermions,  the following CS terms are generated along the RG flow: 
\begin{align}
k_{eff}^{U(1)_1} &=6k -(F-\bar{F}) =0 \\
k_{eff}^{U(1)_1,U(1)_2} &= 2k-(F-\bar{F}) =-2 \\
k_{eff}^{SU(2)} &= k+\frac{1}{2} (F-\bar{F}) =2.
\end{align}
Since the Chern-Simons level for the unbroken $SU(2)$ subgroup is generated, the low-energy pure $SU(2)$ CS theory has a supersymmetric vacuum \cite{Intriligator:2013lca} (non-zero Witten index). This is a necessary condition for this flat direction to be stable and to quantum-mechanically exist. As another necessary condition, the $U(1)_1$ Chern-Simons term correctly vanishes since a non-zero $k_{eff}^{U(1)_1} $ turns on a mass for the vector multiplet. Notice that the mass term contradicts the fact that we are studying the Coulomb flat direction. Due to the non-zero mixed CS term $k_{eff}^{U(1)_1,U(1)_2}$, the bare operator $Y_{SU(2)}^{bare}$ obtains a non-zero $U(1)_2$ charge \cite{Affleck:1982as, Intriligator:2013lca}. Therefore, we need to define a dressed monopole. In terms of the fundamental monopoles, the bare and dressed operators are defined as 
\begin{align}
Y_{SU(2)}^{bare} &:= \sqrt{Y_1 Y_2^2 Y_3^2} \\
Y_d &:= Y_{SU(2)}^{bare}\mathbf{1}_{0,-1} \sim Y_{SU(2)}^{bare} \tilde{Q},
\end{align}
Notice that the bare operator needs a square root since the above monopole has a unbroken non-abelian factor and the minimal magnetic charge becomes smaller than the naive one \cite{Csaki:2014cwa, Amariti:2015kha, Nii:2018bgf, Weinberg:2012pjx, Preskill:1984gd}. The low-energy theory is described by three composite fields $M, B$ and $Y_d$ with a cubic superpotential
\begin{align}
W_{eff} = BMY_d. 
\end{align}
The $F$-flatness condition of $Y_d$ correctly reduces the independent components of the Higgs branch coordinates.

Finally, we give another interpretation of the dressed monopole operator defined above. The bare Coulomb branch operator $\tilde{Y}_{SU(2)}^{bare}$ now induces the following breaking
\begin{align}
SU(4) & \rightarrow SU(2) \times U(1)_1 \times U(1)_2  \\
 {\tiny \yng(1)} & \rightarrow  {\tiny \yng(1)}_{ \,0,-1} + \mathbf{1}_{1,1} +\mathbf{1}_{-1,1} \\
 {\tiny \overline{\yng(1)}} & \rightarrow  {\tiny \yng(1)}_{ \,0,1} +\mathbf{1}_{-1,-1} +\mathbf{1}_{1,-1} \\
 \mathbf{adj.}   & \rightarrow  \mathbf{adj.}_{0,0}+\mathbf{1}_{0,0}+ \mathbf{1}_{0,0}+\mathbf{1}_{2,0}+ \mathbf{1}_{-2,0}  \nonumber \\
 & \qquad \qquad + {\tiny \yng(1)}_{ \,-1,-2} + {\tiny \yng(1)}_{ \,1,2} + {\tiny \yng(1)}_{ \,1,-2}+ {\tiny \yng(1)}_{ \,-1,2},
\end{align}
where the Coulomb branch corresponds to the $U(1)_1$ generator. Notice that this breaking is different from the previous one. Especially, the low-energy $SU(2)$ gauge theory includes massless dynamical quarks and the vacuum of the $SU(2)$ sector is supersymmetric. However, we obtain the effective CS terms for the abelian subgroups 
\begin{align}
k_{eff}^{U(1)_1,U(1)_2} =3,~~~~~~k_{eff}^{U(1)_1}=1.
\end{align}
Due to the CS term $k_{eff}^{U(1)_1}$, the flat direction of this Coulomb branch becomes massive and cannot be a part of the quantum moduli space. However, we can define the dressed monopole operator as follows:
\begin{align}
\tilde{Y}_{SU(2)}^{bare}  {\tiny \yng(1)}_{ \,1,2}   {\tiny \yng(1)}_{ \,0,1} \sim  \tilde{Y}_{SU(2)}^{bare} W_\alpha \tilde{Q},
\end{align}
where ${\tiny \yng(1)}_{ \,1,2} $ comes from the gaugino field whose spin is transmuted into boson statistics on the monopole background \cite{Polyakov:1988md, Dimofte:2011py, Aharony:2015pla}. The quantum numbers of this dressed operator lead to the identification $Y_d \sim \tilde{Y}_{SU(2)}^{bare} W_\alpha \tilde{Q}$. The latter construction of the dressed monopole is consistent with the SCI calculation although it includes massive excitations.

\begin{table}[H]\caption{The 3d $\mathcal{N}=2$ $SU(4)_{k=\frac{1}{2}}$ gauge theory with $(F,\bar{F})=(5,2)$} 
\begin{center}
  \begin{tabular}{|c||c|c|c|c|c|c| } \hline
  &$SU(4)_{k=1/2}$&$SU(5)$&$SU(2)$&$U(1)$&$U(1)$&$U(1)_R$ \\ \hline
$Q$& ${\tiny \yng(1)}$&$\tiny \yng(1)$&1&$1$&0&$0$ \\  
$\tilde{Q}$& ${\tiny \overline{\yng(1)}}$&1&$\tiny \yng(1)$&0&1&0  \\ \hline
$M:=Q\tilde{Q}$&1&${\tiny \yng(1)}$&$\tiny \yng(1)$&$1$&1&$0$ \\ 
$B:=Q^4$&1&${\tiny \overline{\yng(1)}}$&1&4&0&0  \\ \hline
$Y_{SU(2)}^{bare}$&$U(1)_1$: $0$, $U(1)_2:$ $1$&1&1&$-5$&$-2$&$2$ \\
$Y_d:=Y^{bare}_{SU(2)} \tilde{Q}$&1&1&$\tiny \yng(1)$&$-5$&$-1$&$2$  \\ \hline
  \end{tabular}
  \end{center}\label{SU4,52,hk}
\end{table}

\subsection{$SU(4)_{k=1}$ with $(F, \bar{F})=(5,1)$}
The next $SU(4)$ example is a 3d $\mathcal{N}=2$ $SU(4)$ gauge theory with $(F, \bar{F})=(5,1)$ (anti-)fundamental matter fields. The bare Chern-Simons level is set to be $k=1$. The Higgs branch is described by $M:=Q \tilde{Q}$ and $B:=Q^4$. In what follows, we will focus on the Coulomb branch. 

Both because the bare Chern-Simons term is introduced and because the chiral matter content also generates non-zero CS terms along the Coulomb flat directions, almost all the classical Coulomb branches become massive and cannot be a part of the moduli space. We argue that a one-dimensional flat direction $Y_{SU(3)}$ can survive all the quantum corrections. When the bare operator $Y_{SU(3)}$ obtains a non-zero expectation value, the gauge group is spontaneously broken as
\begin{align}
SU(4) & \rightarrow SU(3) \times U(1),~~~t_{U(1)}=\mathrm{diag.} (1,1,1,-3) \\
{\tiny \yng(1)} & \rightarrow {\tiny \yng(1)}_{\, 1} +\mathbf{1}_{-3}     \\
{\tiny \overline{\yng(1)}} & \rightarrow  {\tiny \overline{\yng(1)}}_{\, -1} +\mathbf{1}_{3}   \\
\mathbf{adj.}   & \rightarrow \mathbf{adj.}_0+\mathbf{1}_{0} +  {\tiny \yng(1)}_{\,4} +{\tiny \overline{\yng(1)}}_{\, -4},
\end{align}
where the Coulomb branch corresponds to the $U(1)$ generator. Along the non-zero $Y_{SU(3)}$ direction, the components charged under the $U(1)_1$ subgroup are massive. By integrating out these massive fields at one-loop level, we obtain the following effective Chern-Simons terms
\begin{align}
k_{eff}^{U(1)} &= 12k -3(F-\bar{F}) =0 \\
k_{eff}^{SU(3)} &=k +\frac{1}{2} (F -\bar{F}) = 3,
\end{align}
where $k_{eff}^{U(1)}$ is a Chern-Simons level for the $U(1)_1$ subgroup and it is vanishing. This is consistent with the claim that $Y_{SU(3)}$ becomes a massless flat direction. We should also care about the low-energy dynamics of the pure $SU(3)$ Chern-Simons theory whose CS level is given by $k_{eff}^{SU(3)}$. By computing the Witten index, we can see that the $SU(3)$ sector also has a supersymmetric vacuum. In this way, the flat direction $Y_{SU(3)}$ is stable against all the quantum effects from $SU(3)$ and $U(1)$. Since the unbroken subgroup includes a non-abelian factor, we can define the non-abelian monopole whose magnetic charge is smaller than the naive one \cite{Csaki:2014cwa, Amariti:2015kha, Nii:2018bgf, Weinberg:2012pjx, Preskill:1984gd}. By using the fundamental monopoles $Y_i~(i=1,2,3)$, the minimal monopole operator is given by
\begin{align}
Y_{SU(3)} := (Y_1 Y_2^2 Y_3^3 )^{\frac{1}{3}}, \label{SU42nd}
\end{align}
whose quantum numbers are summarized in Table \ref{SU4,51,1k}. This monopole is associated with a linear combination of the generators $\mathbb{Z}_3 \times U(1) \subset SU(3) \times U(1)$, which explains the fractional power in \eqref{SU42nd}.  Based on this Coulomb branch, the low-energy dynamics is described by $M,B,Y_{SU(3)}$ and an effective superpotential
\begin{align}
W_{eff} = Y_{SU(3)} MB.
\end{align}

From the observation of the superconformal indices, the Coulomb branch discussed above is interpreted in a different way: From the viewpoint of the SCI, which is the summation over the states with all the GNO charges, the bare Coulomb branch corresponds to the following breaking
\begin{align}
SU(4) & \rightarrow SU(2) \times U(1)_1 \times U(1)_2 \\
{\tiny \yng(1)} & \rightarrow {\tiny \yng(1)}_{\, 0,-1}+\mathbf{1}_{1,1} +\mathbf{1}_{-1,1}  \\
{\tiny \overline{\yng(1)}} & \rightarrow  {\tiny \yng(1)}_{\, 0,1} +\mathbf{1}_{1,1} +\mathbf{1}_{1,-1} \\
\mathbf{adj.}   & \rightarrow \mathbf{adj.}_{0,0}+ \mathbf{1}_{0,0} + \mathbf{1}_{0,0} + \mathbf{1}_{2,0} + \mathbf{1}_{-2,0} \nonumber \\
&\qquad + {\tiny \yng(1)}_{\, -1,-2}+{\tiny \yng(1)}_{\, 1,-2}+{\tiny \yng(1)}_{\, 1,2}+{\tiny \yng(1)}_{\, -1,2}.
\end{align}
Along this direction, the massive components, which are charged under the $U(1)_1$ subgroup, are integrated out. This results in the following effective Chern-Simons term:
\begin{align}
k_{eff}^{U(1)_1} =2k=2,~~~~k_{eff}^{U(1)_1,U(1)_2}=F-\bar{F}=4,~~~~k_{eff}^{SU(2)}=k=1  
\end{align}
Since the abelian Chern-Simons levels are non-vanishing, the bare monopole is not-gauge invariant. Hence, we have to define the following dressed operators
\begin{align}
Y_{SU(2)}^{bare} &:=Y_1 Y_2 Y_3 \\
Y_{SU(2)}^{dressed} &:=  Y_{SU(2)}^{bare}  {\tiny \yng(1)}_{\, 1,2} {\tiny \yng(1)}_{\, 1,2} \sim Y_{SU(2)}^{bare} W_\alpha^2 \sim Y_{SU(3)} 
\end{align}
In the expansion of superconformal indices, we can see that the bare Coulomb branch $Y_{SU(2)}^{bare}$ is dressed by two gauginos ${\tiny \yng(1)}_{\, 1,2}$. This operator is bosonic since the components ${\tiny \yng(1)}_{\, 1,2}$ obtain anomalous spins under the monopole background \cite{Polyakov:1988md, Dimofte:2011py, Aharony:2015pla}. This interpretation is completely fine from the index point of view. However, since the low-energy $SU(2)$ gauge theory becomes a pure $SU(2)_{k=1}$ Chern-Simons without matter, its vacuum is non-supersymmetric. Therefore, this flat direction is eliminated from the quantum moduli space. As we discussed above, this Coulomb branch should be interpreted as the flat direction which breaks the gauge group as $SU(3) \times U(1)$.

\begin{table}[H]\caption{The 3d $\mathcal{N}=2$ $SU(4)_{k=1}$ gauge theory with $(F,\bar{F})=(5,1)$} 
\begin{center}
  \begin{tabular}{|c||c|c|c|c|c| } \hline
  &$SU(4)_{k=1}$&$SU(5)$&$U(1)$&$U(1)$&$U(1)_R$ \\ \hline
$Q$& ${\tiny \yng(1)}$&$\tiny \yng(1)$&$1$&0&$0$ \\  
$\tilde{Q}$& ${\tiny \overline{\yng(1)}}$&1&0&1&0  \\ \hline
$M:=Q\tilde{Q}$&1&${\tiny \yng(1)}$&1&1&$0$ \\ 
$B:=Q^4$&1&${\tiny \overline{\yng(1)}}$&4&0&0  \\ \hline
$Y_{SU(3)}$&1&1&$-5$&$-1$&$2$ \\ \hline
  \end{tabular}
  \end{center}\label{SU4,51,1k}
\end{table}

\section{$SU(5)$ gauge group}
In this section, we examine the $SU(5)_k$ example, focusing on the s-confinement phases with a dressed monopole operator. 

\subsection{$SU(5)_{k=\frac{1}{2}}$ with $(F, \bar{F})=(6,3)$}
The s-confinement with a smallest (non-zero) CS level appears in the $SU(5)_{k=\frac{1}{2}}$ CS theory with $(F, \bar{F})=(6,3)$ (anti-)fundamental chiral multiplets. The s-confinement with no CS level was studied in \cite{Aharony:1997bx, Aharony:2013dha, Nii:2018bgf} and we don't here discuss it. The quantum numbers of the elementary fields are summarized in Table \ref{SU5,63,hk}. The Higgs branch is described by a meson $M:=Q \tilde{Q}$ and a baryon $B:=Q^5$. By giving a vev to the baryon, the gauge group is completely broken. Therefore, the dimension of the Higgs branch is $45-24=21$ while the total number of the Higgs branch coordinates is $24$. Then, we expect that there are three constraints between them. As we will see below, the $F$-term flatness condition of the dressed monopole correctly reduces the number of the Higgs branch operators.    

As in the vector-like theory, almost all the classical Coulomb branches are quantum-mechanically lifted since the tree-level CS term serves as a topological mass term for the vector multiplet. Furthermore, the chiral matter content generates effective CS terms along the Coulomb branch. As a result, only a single Coulomb branch operator is non-perturbatively available in the present setup.  When the bare Coulomb branch, denoted by $Y_{SU(2) \times SU(2)}^{bare}$, obtains a non-zero expectation value, the gauge group is broken as
\begin{align}
SU(5) &\rightarrow SU(2) \times SU(2) \times U(1)_1 \times U(1)_2  \label{SU(5)1}\\
{\tiny \yng(1)} & \rightarrow ({\tiny \yng(1)},\cdot)_{\, 1,0}+(\cdot, {\tiny \yng(1)})_{\, 0,1} +(\cdot, \cdot)_{-2,-2} \\
{\tiny \overline{\yng(1)}} & \rightarrow  ({\tiny \yng(1)},\cdot)_{\, -1,0}+(\cdot, {\tiny \yng(1)})_{\, 0,-1} +(\cdot, \cdot)_{2,2} \\
\mathbf{adj.}   & \rightarrow  (\mathbf{adj.} ,\cdot)_{0,0} +(\cdot, \mathbf{adj.} )_{0,0}+(\cdot, \cdot)_{0,0}+(\cdot, \cdot)_{0,0}+(\cdot,{\tiny \yng(1)})_{2,3}+(\cdot, {\tiny \yng(1)})_{-2,-3} \nonumber \\
& \qquad \qquad + ({\tiny \yng(1)}, {\tiny \yng(1)})_{1,-1}+({\tiny \yng(1)},{\tiny \yng(1)})_{-1,1}+({\tiny \yng(1)},\cdot)_{3,2}+({\tiny \yng(1)},\cdot)_{-3,-2},
\end{align} 
where the two $U(1)$ generators are defined as $t_{U(1)_1} := \mathrm{diag.}(1,1,0,0,-2)$ and  $
t_{U(1)_2} :=  \mathrm{diag.} (0,0,1,1,-2)$, respectively. Notice that this breaking pattern is very asymmetric compared to the Coulomb branch in the SQCD theory without a CS term \cite{Aharony:1997bx, Aharony:2013dha, Nii:2018bgf}. Especially, this breaking is not induced by $\braket{\phi_{adj.}}=\mathrm{diag.}(v,v,0,-v,-v)$ although the remaining gauge symmetries are the same. As we will see below, this is important for this (classical) flat direction to survive all the quantum corrections. The bare Coulomb branch here corresponds to the $U(1)_1$ subgroup and its coordinate can be obtained by dualizing the $U(1)_1$ vector superfield into a chiral superfield.

Along the Coulomb branch \eqref{SU(5)1}, the components charged under the $U(1)_1$ subgroup are massive and integrated out. From the $1$-loop diagrams of the massive fermions, the following Chern-Simons terms are generated:
\begin{align}
k_{eff}^{U(1)_1} &= 6k -(F-\bar{F})=0 \\
k_{eff}^{U(1)_1,U(1)_2} &= 4k -2(F-\bar{F}) =-4 \\
k_{eff}^{SU(2)_1} &=k+ \frac{1}{2} (F-\bar{F}) =2 \\
k_{eff}^{SU(2)_2} &=k =\frac{1}{2}.
\end{align}
First of all, the $U(1)_1$ effective CS term $k_{eff}^{U(1)_1}$ vanishes and this is consistent with the fact that we are studying the flat direction associated with the $U(1)_1$ vector multiplet. Namely, no mass is generated to this $U(1)_1$ vector superfield. For the $SU(2)_1$ subgroup, the low-energy theory becomes a pure CS theory without matter and the effective CS term $k_{eff}^{SU(2)_1}$ guarantees that there is a supersymmetric vacuum. For the $SU(2)_2$ subgroup, the low-energy theory has massless quarks $(\cdot, {\tiny \yng(1)})_{\, 0,1}$ and $(\cdot, {\tiny \yng(1)})_{\, 0,-1}$ and there is no constraint on $k_{eff}^{SU(2)_2}$. Since the mixed CS term $k_{eff}^{U(1)_1,U(1)_2}$ is generated, the bare operator $Y_{SU(2) \times SU(2)}^{bare}$ obtains a non-zero $U(1)_2$ charge \cite{Intriligator:2013lca}. Therefore, we need to define a dressed monopole operator. In terms of the fundamental monopoles, the dressed Coulomb branch is defined as
\begin{align}
Y_{SU(2) \times SU(2)}^{bare} &:=\sqrt{Y_1 Y_2^2Y_3^2Y_4^2} \\
Y_d &:= Y_{SU(2) \times SU(2)}^{bare}  ((\cdot, {\tiny \yng(1)})_{\, 0,-1})^2 \sim  Y_{SU(2) \times SU(2)}^{bare}  \tilde{Q}^2,
\end{align}
where the monopole $Y_{SU(2) \times SU(2)}^{bare} $ is a non-abelian monopole whose $U(1)$ gauge symmetry is a linear combination between the $U(1)_1$ and the center symmetry of the second $SU(2)$ \cite{Csaki:2014cwa, Amariti:2015kha}. The flavor indices of $\tilde{Q}^2$ are anti-symmetrized.

Here, we give an alternative interpretation for the above monopole in a way that the SCI naturally captures this operator \cite{Aharony:2015pla}. The bare Coulomb branch $Y_{SU(3)}^{bare}$ induces the gauge symmetry breaking
\begin{align}
SU(5) & \rightarrow SU(3) \times U(1)_1 \times U(1)_2 \\
{\tiny \yng(1)} & \rightarrow {\tiny \yng(1)}_{0,-2}+\mathbf{1}_{1,3}+\mathbf{1}_{-1,3} \\
{\tiny \overline{\yng(1)}} & \rightarrow {\tiny \overline{\yng(1)}}_{0,2}+\mathbf{1}_{-1,-3}+\mathbf{1}_{1,-3}  \\
\mathbf{adj.}   & \rightarrow \mathbf{adj.}_{0,0}+\mathbf{1}_{0,0}+\mathbf{1}_{0,0}+\mathbf{1}_{2,0}+\mathbf{1}_{-2,0} \nonumber \\
& \qquad \qquad + {\tiny \yng(1)}_{-1,-5}+{\tiny \yng(1)}_{1,-5}+{\tiny \overline{\yng(1)}}_{1,5}+{\tiny \overline{\yng(1)}}_{-1,5},
\end{align}
where the Coulomb branch corresponds to the $U(1)_1$ generator. This (classical) flat direction was considered in a vector-like SQCD \cite{Aharony:1997bx} and may be a part of the quantum moduli space. However, in the present situation, the following effective CS terms are generated 
\begin{align}
k_{eff}^{U(1)_1} &= 2k =1  \\
k_{eff}^{U(1)_1,U(1)_2} &=3(F-\bar{F})=9 \\
k_{eff}^{SU(3)} &=k =\frac{1}{2}, 
\end{align}
where $k_{eff}^{U(1)_1}$ and $k_{eff}^{SU(3)}$ solely come from the bare CS term. Since the $U(1)_1$ CS term $k_{eff}^{U(1)_1} $ is non-zero and acts as a topological mass for the $U(1)_1$ vector multiplet, this Coulomb branch, which is classically flat, becomes massive. Therefore, this direction is removed from the quantum moduli space. However, we can formally define the gauge-invariant operator by dressing the bare monopole $Y_{SU(3)}^{bare}$ with a massive gaugino as follows:
\begin{align}
Y_{SU(3)}^{bare} &:= Y_1 Y_2 Y_3 Y_4 \\
Y_d  &\sim Y_{SU(3)}^{bare} ( {\tiny \overline{\yng(1)}}_{0,2})^2 {\tiny \overline{\yng(1)}}_{1,5} \sim Y_{SU(3)}^{bare}  \tilde{Q}^2 W_\alpha.
\end{align}
We can see that this dressed operator has the same quantum numbers as $Y_d $. This viewpoint of the dressed monopole is consistent with the SCI calculation since $Y_{SU(3)}^{bare}$ is associated with the GNO charge. By using the monopole operator $Y_d$ defined above, the low-energy effective superpotential becomes
\begin{align}
W_{eff}=Y_d  MB,
\end{align}
whose $F$-term condition for $Y_d$ removes three components from the Higgs branch coordinates and correctly explains the dimensions of the Higgs moduli space.

\begin{table}[H]\caption{The 3d $\mathcal{N}=2$ $SU(5)_{k=\frac{1}{2}}$ gauge theory with $(F,\bar{F})=(6,3)$} 
\begin{center}
  \begin{tabular}{|c||c|c|c|c|c|c| } \hline
  &$SU(5)_{k=1/2}$&$SU(6)$&$SU(3)$&$U(1)$&$U(1)$&$U(1)_R$ \\ \hline
$Q$& ${\tiny \yng(1)}$&$\tiny \yng(1)$&1&$1$&0&$0$ \\  
$\tilde{Q}$& ${\tiny \overline{\yng(1)}}$&1&$\tiny \yng(1)$&0&1&0  \\ \hline
$M:=Q\tilde{Q}$&1&${\tiny \yng(1)}$&$\tiny \yng(1)$&$1$&1&$0$ \\ 
$B:=Q^5$&1&${\tiny \overline{\yng(1)}}$&1&5&0&0  \\ \hline
$Y_{SU(2) \times SU(2)}^{bare}$&$U(1)_2$: $2$&1&1&$-6$&$-3$&$2$  \\
$Y_d:=Y_{SU(2) \times SU(2)}^{bare} \tilde{Q}^2$&1&1&${\tiny \overline{\yng(1)}}$&$-6$&$-1$&$2$  \\ \hline
  \end{tabular}
  \end{center}\label{SU5,63,hk}
\end{table}

\subsection{$SU(5)_{k=1}$ with $(F, \bar{F})=(6,2)$}
The second example is a 3d $\mathcal{N}=2$ $SU(5)_{k=1}$ gauge theory with $(F, \bar{F})=(6,2)$ (anti-)fundamental matter fields, which is obtained from the previous theory (Table \ref{SU5,63,hk}) via a real mass deformation for a single anti-fundamental quark $\tilde{Q}_{\bar{f}=3}$. Therefore, we can expect that the theory again exhibits s-confinement. On the magnetic side, the real mass makes $M_{i,3}$ and $Y_d^{3}$ massive. Therefore, this theory also allows a dressed monopole even if the UV theory has a non-zero CS term. The Higgs branch is almost the same as the previous one with reduction of $\bar{F}$ so we will focus on the Coulomb branch below.

When the bare Coulomb branch, denoted by $Y_{SU(3)}^{bare}$, obtains a non-zero expectation value, the gauge group is spontaneously broken as
\begin{align}
SU(5) & \rightarrow SU(3) \times U(1)_1 \times U(1)_2  \label{SU562CB}\\
{\tiny \yng(1)} & \rightarrow  {\tiny \yng(1)}_{\,1,0} +\mathbf{1}_{0,1}+\mathbf{1}_{-3,-1} \\
{\tiny \overline{\yng(1)}} & \rightarrow {\tiny \overline{\yng(1)}}_{\,-1,0} +\mathbf{1}_{0,-1}+\mathbf{1}_{3,1} \\
\mathbf{adj.}   & \rightarrow \mathbf{adj.}_{0,0}+\mathbf{1}_{0,0}+\mathbf{1}_{0,0} +\mathbf{1}_{3,2}+\mathbf{1}_{-3,-2} \nonumber \\
& \qquad \qquad + {\tiny \yng(1)}_{\,1,-1}+ {\tiny \yng(1)}_{\,4,1} +{\tiny \overline{\yng(1)}}_{\,-1,1}+{\tiny \overline{\yng(1)}}_{\,-4,-1},
\end{align}
where the two $U(1)$ generators are defined as $t_{U(1)_1} = \mathrm{diag.} (1,1,1,0,-3)$ and $t_{U(1)_2}=\mathrm{diag.} (0,0,0,1,-1)$, respectively. The Coulomb branch corresponds to the first $U(1)_1$ generator $t_{U(1)_1}$ and is dual to the $U(1)_1$ vector superfield. By integrating out all the massive components which are charged under $U(1)_1$, we obtain the effective CS terms
\begin{align}
k_{eff}^{U(1)_1} &= 12k -3(F-\bar{F}) = 0\\
k_{eff}^{U(1)_1,U(1)_2} &= 3k -\frac{3}{2}(F-\bar{F}) =-3 \\
k_{eff}^{SU(3)} &= k+\frac{1}{2}(F-\bar{F}) =3.
\end{align}
We can see that the $U(1)_1$ CS term $k_{eff}^{U(1)_1}$ correctly cancels out and this direction is quantum-mechanically flat. For the $SU(3)$ sector, the low-energy limit becomes a pure $SU(3)$ CS theory without matter. Since $|k_{eff}^{SU(3)}| \ge 3$, the low-energy $SU(3)$ theory has a supersymmetric vacuum (non-zero Witten index) \cite{Intriligator:2013lca}. Due to the mixed CS term $k_{eff}^{U(1)_1,U(1)_2} $ between $U(1)_1$ and $U(1)_2$, the bare operator $Y_{SU(3)}^{bare}$ has a non-zero $U(1)_2$ charge. As a result, we need to introduce a dressed operator
\begin{align}
Y_{SU(3)}^{bare} &:= \left( Y_1 Y_2^2 Y_3^3 Y_4^3 \right)^{\frac{1}{3}} \\
Y_d &:= Y_{SU(3)}^{bare}   \mathbf{1}_{0,-1} \sim  Y_{SU(3)}^{bare}  \tilde{Q}.
\end{align}
In terms of this monopole, the effective superpotential becomes
\begin{align}
W_{eff} = Y_d MB.
\end{align}
It is straightforward to derive this superpotential from the previous result via a real mass deformation.

For completeness of our study, we here give an alternative picture of the Coulomb branch, which is consistent with the SCI calcualtion. When the bare Coulomb branch, denoted by ${Y'}_{SU(3)}^{bare}$, obtains a non-zero vev, the gauge group is spontaneously broken to
\begin{align}
SU(5) & \rightarrow SU(3) \times U(1)_1 \times U(1)_2 \\
{\tiny \yng(1)} & \rightarrow {\tiny \yng(1)}_{\, 0,-2}+\mathbf{1}_{1,3}+\mathbf{1}_{-1,3} \\
{\tiny \overline{\yng(1)}} & \rightarrow {\tiny \overline{\yng(1)}}_{\,0,2}+\mathbf{1}_{-1,-3}+\mathbf{1}_{1,-3}  \\
\mathbf{adj.}   & \rightarrow \mathbf{adj.}_{0,0}+\mathbf{1}_{0,0}+\mathbf{1}_{0,0}+\mathbf{1}_{2,0}+\mathbf{1}_{-2,0} \nonumber \\
& \qquad + {\tiny \yng(1)}_{\, -1,-5}+{\tiny \yng(1)}_{\, 1,-5}+{\tiny \overline{\yng(1)}}_{\, 1,5}+{\tiny \overline{\yng(1)}}_{\, -1,5},
\end{align}
where the Coulomb branch corresponds to the $U(1)_1$ subgroup and $\mathbf{adj.}$ corresponds to the gaugino field. Notice that this breaking is similar to \eqref{SU562CB} but the $U(1)$ generators are different. Along the Coulomb branch, the components charged under the $U(1)_1$ subgroup are massive and lead to the effective CS terms
\begin{align}
k_{eff}^{U(1)_1} &= 2k=2 \\
k_{eff}^{U(1)_1,U(1)_2} &= 3(F-\bar{F}) =12 \\
k_{eff}^{SU(3)} &=k =1
\end{align}
Due to the $U(1)_1$ CS term $k_{eff}^{U(1)_1}$, this flat direction becomes massive. In order to cancel the $U(1)_2$ charge of ${Y'}_{SU(3)}^{bare}$ from the mixed CS term $k_{eff}^{U(1)_1,U(1)_2}$, we need to define the dressed monopole by using the massive gauginos as follows:
\begin{align}
Y_d &\sim {Y'}_{SU(3)}^{bare} {\tiny \overline{\yng(1)}}_{\,0,2} ({\tiny \overline{\yng(1)}}_{\, 1,5})^2\sim {Y'}_{SU(3)}^{bare} \tilde{Q} W_\alpha^2. 
\end{align}
From the comparison of the quantum numbers between $Y_d$ and ${Y'}_{SU(3)}^{bare} \tilde{Q} W_\alpha^2$, we can identify them.

\begin{table}[H]\caption{The 3d $\mathcal{N}=2$ $SU(5)_{k=1}$ gauge theory with $(F,\bar{F})=(6,2)$} 
\begin{center}
  \begin{tabular}{|c||c|c|c|c|c|c| } \hline
  &$SU(5)_{k=1}$&$SU(6)$&$SU(2)$&$U(1)$&$U(1)$&$U(1)_R$ \\ \hline
$Q$& ${\tiny \yng(1)}$&$\tiny \yng(1)$&1&$1$&0&$0$ \\  
$\tilde{Q}$& ${\tiny \overline{\yng(1)}}$&1&$\tiny \yng(1)$&0&1&0  \\ \hline
$M:=Q\tilde{Q}$&1&${\tiny \yng(1)}$&$\tiny \yng(1)$&$1$&1&$0$ \\ 
$B:=Q^5$&1&${\tiny \overline{\yng(1)}}$&1&5&0&0  \\ \hline
$Y_{SU(3)}^{bare}$&$U(1)_2$: $1$&1&1&$-6$&$-2$&$2$  \\
$Y_d:=Y_{SU(3)}^{bare} \tilde{Q}$&1&1&${\tiny \yng(1)}$&$-6$&$-1$&$2$  \\ \hline
  \end{tabular}
  \end{center}\label{SU5,62,1k}
\end{table}

\subsection{$SU(5)_{k=\frac{3}{2}}$ with $(F, \bar{F})=(6,1)$}
In this subsection, we study the 3d $\mathcal{N}=2$ $SU(5)_{k=\frac{3}{2}}$ gauge theory with six fundamental and one anti-fundamental matter fields. The elementary fields and their quantum numbers are summarized in Table \ref{SU(5),61,3/2k}. This theory can be obtained from the previous theory (Table \ref{SU5,62,1k}) via a real mass deformation for an anti-fundamental quark $\tilde{Q}^{\bar{F}=2}$. On the dual side, the corresponding mass term makes a single component of the dressed monopole massive while the other component remains massless. Therefore, the present theory also has a one-component monopole operator although the breaking pattern of the gauge symmetry along the Coulomb moduli space will be different from the previous example. 

In this example, the bare Coulomb branch, denoted by $Y_{SU(4)}$, leads to the following gauge symmetry breaking 
\begin{align}
SU(5) & \rightarrow SU(4) \times U(1),~~~t_{U(1)}:= \mathrm{diag.} (1,1,1,1,-4) \\
{\tiny \yng(1)} & \rightarrow {\tiny \yng(1)}_{\, 1}+\mathbf{1}_{-4}  \\
{\tiny \overline{\yng(1)}} & \rightarrow {\tiny \overline{\yng(1)}}_{\,-1} +\mathbf{1}_{4}  \\
\mathbf{adj.}   & \rightarrow \mathbf{adj.}_{0}+\mathbf{1}_{0}+ {\tiny \yng(1)}_{\, 5}+{\tiny \overline{\yng(1)}}_{\,-5},
\end{align}
where $Y_{SU(4)}$ is constructed from the $U(1)$ vector superfield. Along the Coulomb branch, the components charged under the $U(1)$ symmetry are massive and integrated out, which results in the following effective CS terms
\begin{align}
k_{eff}^{U(1)} &=20k -6(F-\bar{F}) =0 \\
k_{eff}^{SU(4)} &=k +\frac{1}{2}(F-\bar{F}) =4
\end{align}
We can see that the $U(1)$ CS term $k_{eff}^{U(1)}$ is vanishing and thus the $Y_{SU(4)}$ direction is indeed flat. For the $SU(4)$ part whose low-energy limit becomes a pure CS theory without matter, the effective CS term $k_{eff}^{SU(4)}$ correctly ensures that there is a supersymmetric vacuum \cite{Intriligator:2013lca}. Therefore, this Coulomb branch labeled by $Y_{SU(4)}$ becomes a part of the quantum moduli space. Since there is only a single $U(1)$ factor, this bare operator is gauge-invariant and needs no ``dressing.'' In terms of the fundamental monopoles, this operator is defined as
\begin{align}
Y_{SU(4)} := \left( Y_1 Y_2^2 Y_3^3 Y_4^4 \right)^{\frac{1}{4}}.
\end{align}
Notice that the breaking pattern includes a non-abelian factor and $Y_{SU(4)}$ corresponds to the non-abelian monopole as before \cite{Csaki:2014cwa, Preskill:1984gd, Weinberg:2012pjx}.

The Coulomb branch introduced above can be interpreted in a different way that is consistent with the SCI computation. This is because the SCI is summing up the states with all the GNO charges and there is no state with the magnetic charge of the non-abelian monopole. Therefore, from the SCI viewpoint, the Coulomb branch $Y_{SU(4)}$ appears as a different contribution in a certain GNO sector. The bare Coulomb branch $Y_{SU(3)}^{bare}$ induces the gauge symmetry breaking
\begin{align}
SU(5) & \rightarrow SU(3) \times U(1)_1 \times U(1)_2 \\
{\tiny \yng(1)} & \rightarrow {\tiny \yng(1)}_{\, 0,-2}+\mathbf{1}_{1,3}+\mathbf{1}_{-1,3} \\
{\tiny \overline{\yng(1)}} & \rightarrow {\tiny \overline{\yng(1)}}_{\,0,2}+\mathbf{1}_{-1,-3}+\mathbf{1}_{1,-3}  \\
\mathbf{adj.}   & \rightarrow \mathbf{adj.}_{0,0}+\mathbf{1}_{0,0}+\mathbf{1}_{0,0}+\mathbf{1}_{2,0}+\mathbf{1}_{-2,0} \nonumber \\
& \qquad \qquad + {\tiny \yng(1)}_{\, -1,-5}+{\tiny \yng(1)}_{\, 1,-5}+{\tiny \overline{\yng(1)}}_{\, 1,5}+{\tiny \overline{\yng(1)}}_{\, -1,5},
\end{align}
where the Coulomb branch is associated with the $U(1)_1$ generator and $\mathbf{adj.}$ is a gaugino field. Along the Coulomb branch $Y_{SU(3)}^{bare}$, the massive components are integrated out and generate various effective CS terms
\begin{align}
k_{eff}^{U(1)_1} &= 2k=3 \\
k_{eff}^{U(1)_1,U(1)_2} &= 3(F-\bar{F}) =15 \\
k_{eff}^{SU(3)} &=k =\frac{3}{2}. 
\end{align}
We can see that this flat direction is lifted since the CS term $k_{eff}^{U(1)_1}$ acts as a topological mass term for the Coulomb branch. Therefore, this branch is eliminated but in the SCI computation, the true Coulomb branch $Y_{SU(4)}$ comes from this sector. Since there is a mixed CS term $k_{eff}^{U(1)_1,U(1)_2}$, we need to define the dressed monopole operator
\begin{align}
Y_{SU(4)} \sim Y_{SU(3)}^{bare} ({\tiny \overline{\yng(1)}}_{\, 1,5})^3 \sim Y_{SU(3)}^{bare} W_\alpha^3,
\end{align}
which is dressed by three massive gauginos. Notice that under the monopole background, these gauginos are transmuted into bosons \cite{Polyakov:1988md, Dimofte:2011py, Aharony:2015pla}. The low-energy dynamics is described by three composites $M,B$ and $Y_{SU(4)}$ with a cubic superpotential 
\begin{align}
W_{eff} =Y_{SU(4)} M B,
\end{align}
which is consistent with all the symmetries in Table \ref{SU(5),61,3/2k}.

\begin{table}[H]\caption{The 3d $\mathcal{N}=2$ $SU(5)_{k=\frac{3}{2}}$ gauge theory with $(F,\bar{F})=(6,1)$} 
\begin{center}
  \begin{tabular}{|c||c|c|c|c|c| } \hline
  &$SU(5)_{k=3/2}$&$SU(6)$&$U(1)$&$U(1)$&$U(1)_R$ \\ \hline
$Q$& ${\tiny \yng(1)}$&$\tiny \yng(1)$&$1$&0&$0$ \\  
$\tilde{Q}$& ${\tiny \overline{\yng(1)}}$&1&0&1&0  \\ \hline
$M:=Q\tilde{Q}$&1&${\tiny \yng(1)}$&$1$&1&$0$ \\ 
$B:=Q^5$&1&${\tiny \overline{\yng(1)}}$&5&0&0  \\ \hline
$Y_{SU(4)}$&1&1&$-6$&$-1$&$2$  \\ \hline
  \end{tabular}
  \end{center}\label{SU(5),61,3/2k}
\end{table}

\subsection{$SU(5)_{k=\frac{1}{2}}$ with $(F, \bar{F})=(6,1)$}
The final s-confining example, where the Coulomb branch survives all the quantum corrections, is a 3d $\mathcal{N}=2$ $SU(5)_{k=\frac{1}{2}}$ gauge theory with six fundamental and one anti-fundamental chiral multiplets. This theory comes from the $SU(5)_{k=0}$ SQCD theory with $(F, \bar{F})=(6,2)$ (anti-)fundamental matter fields \cite{Nii:2018bgf} via a real mass deformation to $\tilde{Q}^{\bar{F}=2}$. Therefore, the present theory will also exhibit s-confinement. The Higgs branch is described by a meson $M:=Q \tilde{Q}$ and a baryon $B:=Q^5$. The bare Coulomb branch, which is denoted by $Y_{SU(3) \times SU(2)}$, induces the following gauge symmetry breaking
\begin{align}
SU(5) & \rightarrow SU(3) \times SU(2) \times U(1) \\
{\tiny \yng(1)} & \rightarrow ({\tiny \yng(1)},\cdot)_{2}+(\cdot, {\tiny \yng(1)})_{-3} \\
{\tiny \overline{\yng(1)}} & \rightarrow  ({\tiny \overline{\yng(1)}} , \cdot)_{-2} +(\cdot,{\tiny \yng(1)} )_3 \\
\mathbf{adj.}   & \rightarrow (\mathbf{adj.} , \cdot)_{0}+(\cdot,\mathbf{adj.} )_0+ (\cdot,\cdot)_0 +({\tiny \yng(1)} ,{\tiny \yng(1)} )_{5}+({\tiny \overline{\yng(1)}} ,{\tiny \yng(1)} )_{-5},
\end{align}
where the $U(1)$ generator is defined by $t_{U(1)}=\mathrm{diag.} (2,2,2,-3,-3)$ and the chiral superfield $Y_{SU(3) \times SU(2)}$ is constructed by dualizing the $U(1)$ vector superfield. Along the Coulomb branch, the components charged under the $U(1)$ subgroup are massive, which are integrated out and generate various effective CS levels
\begin{align}
k_{eff}^{U(1)} &=30k -3(F-\bar{F}) =0\\
k_{eff}^{SU(3)} &=k+\frac{1}{2} (F -\bar{F}) =3 \\
k_{eff}^{SU(2)} &=k-\frac{1}{2} (F -\bar{F}) =-2.
\end{align}
The CS term for the $U(1)$ subgroup correctly vanishes and the moduli field $Y_{SU(3) \times SU(2)}$ can be exactly flat. For the $SU(3) \times SU(2)$ part, the low-energy theory becomes a pure CS theory and these CS levels correctly lead to the existence of a supersymmetric vacuum. Since there is a single $U(1)$ subgroup, the bare Coulomb branch does not need ``dressing.''
In terms of the fundamental monopoles, the Coulomb branch operator is expressed as
\begin{align}
Y_{SU(3) \times SU(2)} := \left( Y_1^2 Y_2^4Y_3^6 Y_4^3  \right)^{\frac{1}{6}},
\end{align}
where the fractional power above indicates that this monopole is a non-abelian monopole whose magnetic charge is defined by combining the $U(1)$ and $\mathbb{Z}_2 \times \mathbb{Z}_3$ center symmetries \cite{Csaki:2014cwa, Preskill:1984gd, Weinberg:2012pjx}. The quantum numbers of this operator are summarized in Table \ref{SU5,61,hk}.

For completeness, we will give an alternative interpretation of the dressed monopole in a way that is consistent with the SCI analysis. The insertion of the bare monopole operator, denoted by $Y^{bare}_{SU(3)}$, induces the gauge symmetry breaking 
\begin{align}
SU(5) & \rightarrow SU(3) \times U(1)_1 \times U(1)_2 \\
{\tiny \yng(1)} & \rightarrow {\tiny \yng(1)}_{\, 0,-2}+\mathbf{1}_{1,3}+\mathbf{1}_{-1,3} \\
{\tiny \overline{\yng(1)}} & \rightarrow {\tiny \overline{\yng(1)}}_{\,0,2}+\mathbf{1}_{-1,-3}+\mathbf{1}_{1,-3}  \\
\mathbf{adj.}   & \rightarrow \mathbf{adj.}_{0,0}+\mathbf{1}_{0,0}+\mathbf{1}_{0,0}+\mathbf{1}_{2,0}+\mathbf{1}_{-2,0} \nonumber \\
& \qquad + {\tiny \yng(1)}_{\, -1,-5}+{\tiny \yng(1)}_{\, 1,-5}+{\tiny \overline{\yng(1)}}_{\, 1,5}+{\tiny \overline{\yng(1)}}_{\, -1,5},
\end{align}
where the Coulomb branch is associated with the $U(1)_1$ generator. $\mathbf{adj.} $ denotes a gaugino field. By integrating out the massive components, we obtain various effective CS terms
\begin{align}
k_{eff}^{U(1)_1} &= 2k=1 \\
k_{eff}^{U(1)_1,U(1)_2} &= 3(F-\bar{F}) =15 \\
k_{eff}^{SU(3)} &=k =\frac{1}{2},
\end{align}
which signals that the $U(1)_1$ flat direction obtains a topological mass term proportional to $k_{eff}^{U(1)_1}$. Due to the mixed CS term $k_{eff}^{U(1)_1,U(1)_2}$, we need to construct a dressed operator consisting of the massive fields  
\begin{align}
Y_{SU(3) \times SU(2)} \sim Y^{bare}_{SU(3)} ({\tiny \overline{\yng(1)}}_{\, 1,5})^2 {\tiny \overline{\yng(1)}}_{\, -1,5} \sim  Y^{bare}_{SU(3)}  W_\alpha^3,
\end{align}
which has the same quantum numbers as $Y_{SU(3) \times SU(2)}$ and it is natural to identify them.

The low-energy dynamics is described by the Higgs branch operators $M, B$ and a bare Coulomb branch $Y_{SU(3) \times SU(2)}$. The effective superpotential can be determined from the symmetries in Table \ref{SU5,61,hk} as
\begin{align}
W_{eff} =Y_{SU(3) \times SU(2)} MB,
\end{align}
which is a cubic superpotential and flows to a non-trivial fixed point. 

\begin{table}[H]\caption{The 3d $\mathcal{N}=2$ $SU(5)_{k=\frac{1}{2}}$ gauge theory with $(F,\bar{F})=(6,1)$} 
\begin{center}
  \begin{tabular}{|c||c|c|c|c|c| } \hline
  &$SU(5)_{k=1/2}$&$SU(6)$&$U(1)$&$U(1)$&$U(1)_R$ \\ \hline
$Q$& ${\tiny \yng(1)}$&$\tiny \yng(1)$&$1$&0&$0$ \\  
$\tilde{Q}$& ${\tiny \overline{\yng(1)}}$&1&0&1&0  \\ \hline
$M:=Q\tilde{Q}$&1&${\tiny \yng(1)}$&$1$&1&$0$ \\ 
$B:=Q^5$&1&${\tiny \overline{\yng(1)}}$&5&0&0  \\ \hline
$Y_{SU(3) \times SU(2)}$&1&1&$-6$&$-1$&$2$  \\ \hline
  \end{tabular}
  \end{center}\label{SU5,61,hk}
\end{table}

\section{Confinement phases for $SU(N)_k$}
Now, we can generalize the previous argument to a generic $SU(N)_k$ gauge group by focusing on confinement phases. In Section 6.1, we construct an s-confinement phase with a dressed monopole, where the dual description is given by gauge-singlet chiral superfields with an effective superpotential. In Section 6.2, we will discuss the confinement phase whose low-energy dynamics possesses a quantum-deformed moduli space where the baryon and Coulomb branch operators are non-perturbatively merged together.

\subsection{$SU(N)_k$ with $(F,\bar{F})=(N+1,N-1-2a-2k)$}
We here discuss an s-confinement phase in a 3d $\mathcal{N}=2$ $SU(N)_k$ gauge theory with $N+1$ fundamental and $N-1-2a-2k$ anti-fundamental matters, where $a$ is a non-negative integer. The CS level $k$ is a positive integer or a half-odd integer, depending on the value of $\frac{1}{2}(F-\bar{F})$. Although this s-confinement phase immediately follows from the chiral $SU(N)_{k=0}$ gauge theory \cite{Nii:2018bgf} via a real mass deformation or from the $SU(N)_k$ CS duality \cite{Aharony:2014uya}, the quantum structure of the Coulomb branch was not explicitly stated there. Then, it would be valuable to explicitly show a detailed analysis of the Coulomb branch that is quantum-mechanically allowed.   

By following the previous analysis of the Coulomb branch, we will consider the Coulomb moduli space where the gauge group is spontaneously broken to $SU(A) \times SU(N-A-B) \times SU(B) \times U(1)_1 \times U(1)_2$. This breaking is generated by a non-zero vev of the adjoint scalar in the $SU(N)$ vector superfield
\begin{align}
\braket{\phi_{adj}} =\mathrm{diag.} (\underbrace{B,\cdots,B}_{A}, \underbrace{0,\cdots,0}_{N-A-B},  \underbrace{-A, \cdots, -A}_{B}),
\end{align}
which is traceless as it should be. 
The Coulomb branch $Y^{bare}$ is associated with the $U(1)_1$ generator and hence the effective CS term $k_{eff}^{U(1)_1}$ must vanish for the flatness of the Coulomb moduli space. Since there are two non-abelian factors $SU(A) \times SU(B)$, whose low-energy dynamics contain no massless dynamical quark, the corresponding CS factors $|k_{eff}^{SU(A)}|$ and $|k_{eff}^{SU(B)}|$ must be greater than the ranks of the two gauge groups. Otherwise, the pure $SU(A) \times SU(B)$ CS theory spontaneously breaks supersymmetry with zero Witten index. For an $SU(N-A-B)$ sector, the low-energy theory has massless dynamical quarks and then no constraint for $A$ and $B$ appears.

From this argument, we find that the following breaking pattern survives all the quantum corrections:
\begin{align}
SU(N) & \rightarrow SU(2k+a+1) \times SU(N-2k -2a-2) \times SU(a+1) \times U(1)_1 \times U(1)_2 \\
{\tiny \yng(1)} & \rightarrow  ({\tiny \yng(1)},\cdot, \cdot)_{a+1,0} +(\cdot, {\tiny \yng(1)},\cdot)_{0,a+1} +(\cdot,\cdot, {\tiny \yng(1)})_{-(2k+a+1),-(N-2k-2a-2)}  \\
{\tiny \overline{\yng(1)}} & \rightarrow ({\tiny \overline{\yng(1)}}, \cdot, \cdot)_{-(a+1),0} +(\cdot, {\tiny \overline{\yng(1)}}, \cdot)_{0, a+1} +(\cdot, \cdot, {\tiny \overline{\yng(1)}})_{2k+a+1,N-2k-2a-2}
\end{align}
The components charged under the $U(1)_1$ symmetry are massive along the Coulomb branch and must be integrated out. This leads to the following effective CS terms for the abelian and non-abelian gauge groups. 
\begin{align}
k_{eff}^{U(1)_1} &=k(a+1)(2k+a+1) \left[(2k+2a+2) -(F-\bar{F})  \right] =0 \\
k_{eff}^{U(1)_1, U(1)_2} &= (a+1)(2k+a+1)(N-2k-2a-2) \left(k-  \frac{1}{2}(F-\bar{F})  \right) \nonumber \\
&=-(a+1)^2 (2k+a+1 ) (N-2k-2a-2) \\
k_{eff}^{SU(2k+a+1)} &= k+ \frac{1}{2}(F-\bar{F})  =2k+a+1 \\
k_{eff}^{SU(a+1)} &=k+\frac{1}{2}(F-\bar{F})  =-a-1
\end{align}
For $(F,\bar{F})=(N,N-1-2a-2k)$, the $U(1)_1$ CS term $k_{eff}^{U(1)_1} $ beautifully cancels out. The non-abelian CS factors imply that the low-energy $SU(2k+a+1) \times SU(a+1)$ pure CS gauge theory has a supersymmetric vacuum \cite{Intriligator:2013lca}. This is consistent with the fact that we are now studying the Coulomb moduli space that by definition preserves supersymmetry. Since the mixed CS term $k_{eff}^{U(1)_1, U(1)_2}$ is non-zero, the bare Coulomb branch obtains a non-zero $U(1)_2$ charge proportional to $k_{eff}^{U(1)_1, U(1)_2}$. Therefore, it is necessary to perform a ``dressing'' procedure as in the previous sections.

In terms of the fundamental monopoles $Y_i \sim \exp(\phi_i -\phi_{i+1})$, which correspond to the fundamental roots, the bare Coulomb branch $Y^{bare}$ considered above is expressed as follows:
\begin{align} 
Y^{bare} & := \exp \left[ \frac{1}{2k+a+1}\phi_1 +\cdots +  \frac{1}{2k+a+1} \phi_{2k+a+1} -\frac{1}{a+1} \phi_{N-a} -\cdots -\frac{1}{a+1} \phi_{N-1} \right] \\
&:= \left( Y_1^{a+1} Y_2^{2(a+1)} \cdots Y_{2k+a+1}^{(2k+a+1)(a+1)} Y_{2k+a+2}^{(2k+a+1)(a+1)} \cdots Y_{N-a-1}^{(2k+a+1)(a+1)} \right.  \nonumber \\ 
&\qquad \qquad  \times   \left. Y_{N-a}^{(2k+a+1)a} Y_{N-a+1}^{(2k+a+1)(a-1)} \cdots Y_{N-1}^{2k+a+1}    \right)^{\frac{1}{(a+1)(2k+a+1)}}  \label{DCBscon}
\end{align}
Notice that the bare operator $Y^{bare}$ associated with a non-abelian monopole and its magnetic charge becomes smaller than the naive one \cite{Csaki:2014cwa, Amariti:2015kha, Nii:2018bgf, Preskill:1984gd}, which explains the fractional power of \eqref{DCBscon}. 
By using this, we can define the gauge-invariant operator
\begin{align} 
Y_d &:= Y^{bare} \left( (\cdot, {\tiny \overline{\yng(1)}}, \cdot )_{0,-(a+1)} \right)^{N-2k-2a-2} \sim Y^{bare} \tilde{Q}^{N-2k-2a-2},
\end{align}
where the color indices of $\tilde{Q}^{N-2k-2a-2}$ are contracted by an epsilon tensor of the $SU(N-2k-2a-2)$ gauge group. The quantum numbers of $Y_d$ are summarized in Table \ref{SU(N)scon}.

For completeness, we also give another interpretation of the dressed monopole, which is consistent with the SCI point of view. The bare Coulomb branch $Y_{SU(N-2)}^{bare}$ corresponds to the breaking $SU(N) \rightarrow SU(N-2) \times U(1)_1 \times U(1)_2$. Due to the (mixed) CS terms between the $U(1)_1$ and $U(1)_2$ subgroups, we need to define a dressed operator
\begin{align}
Y_d &\sim Y_{SU(N-2)}^{bare} ({\tiny \overline{\yng(1)}}_{\,1,N})^{2k} ({\tiny \overline{\yng(1)}}_{\,-1,N}{\tiny \overline{\yng(1)}}_{\,1,N})^{a} ({\tiny \overline{\yng(1)}}_{\,0,2})^{N-2k-2a-2} \\
&\sim Y_{SU(N-2)}^{bare} W_\alpha^{2k} W_\alpha^{2a} \tilde{Q}^{N-2k-2a-2},
\end{align}
where the color indices of $W_\alpha^{2k} W_\alpha^{2a} \tilde{Q}^{N-2k-2a-2}$ are contracted by an epsilon tensor of the unbroken $SU(N-2)$ gauge group. 
The components ${\tiny \overline{\yng(1)}}_{\,1,N}$ and ${\tiny \overline{\yng(1)}}_{\,-1,N}$ are from the gaugino field and they are transmuted into bosons under the monopole background \cite{Polyakov:1988md, Dimofte:2011py}. Since the $U(1)_1$ gauge field has a non-zero CS level, this flat direction is massive and therefore this dressed operator should be interpreted as in \eqref{DCBscon} from the low-energy viewpoint.
The low-energy dynamics is described by the Higgs branch operators $M, B$ and the Coulomb branch $Y_d$ with an effective superpotential
\begin{align}
W_{eff} = BM Y_d,
\end{align}
which is consistent with all the symmetries listed in Table \ref{SU(N)scon}.

\begin{table}[H]\caption{The 3d $\mathcal{N}=2$ $SU(N)_{k}$ gauge theory with $(F,\bar{F})=(N+1,N-1-2a-2k)$} 
\begin{center}\scalebox{0.79}{
  \begin{tabular}{|c||c|c|c|c|c|c| } \hline
  &\footnotesize $SU(N)_{k}$&\footnotesize $SU(N+1)$&\footnotesize $SU(N-1-2a-2k)$&\footnotesize $U(1)$& \footnotesize $U(1)$& \footnotesize $U(1)_R$ \\ \hline
$Q$& ${\tiny \yng(1)}$&$\tiny \yng(1)$&1&$1$&0&$0$ \\  
$\tilde{Q}$& ${\tiny \overline{\yng(1)}}$&1&${\tiny \yng(1)}$&0&1&0  \\ \hline
$M:=Q\tilde{Q}$&1&${\tiny \yng(1)}$&${\tiny \yng(1)}$&$1$&1&$0$ \\ 
$B:=Q^N$&1&${\tiny \overline{\yng(1)}}$&1&$N$&0&0  \\ \hline
$Y^{bare}$& \scriptsize $U(1)_2$: $(a+1)(N-2k-2a-2)$&1&1&$-(N+1)$& \scriptsize $-(N-2a-2k -1)$&2  \\
\scriptsize  $Y_d :=Y^{bare}\tilde{Q}^{N-2k-2a-2}$&1&1&${\tiny \overline{\yng(1)}}$&$-(N+1)$&$-1$&$2$  \\ \hline
  \end{tabular}}
  \end{center}\label{SU(N)scon}
\end{table}

\subsection{$SU(N)_k$ with $(F,\bar{F})=(N,N-2a-2k)$}

By giving a complex mass to one flavor of the previous theory in Table \ref{SU(N)scon} and by integrating out it, we can construct a confining phase with a quantum-deformed moduli space. The UV description becomes a 3d $\mathcal{N}=2$ $SU(N)_k$ gauge theory with $N$ fundamental and $N-2a-2k$ anti-fundamental matter fields, where $a$ is a positive integer. The bare Coulomb branch $Y^{bare}$ induces the gauge symmetry breaking
\begin{align}
SU(N) & \rightarrow SU(2k+a) \times SU(N-2k -2a) \times SU(a) \times U(1)_1 \times U(1)_2 \\
t_{U(1)_1} &:= \mathrm{diag.} (a,\cdots,a, 0, \cdots,0,-(2k+a),\cdots,-(2k+a)) \\
t_{U(1)_2} &:= \mathrm{diag.} (0,\cdot,0,a,\cdots,a,-(N-2k-2a),\cdots,-(N-2k-2a)) \\
{\tiny \yng(1)} & \rightarrow ({\tiny \yng(1)},\cdot,\cdot)_{a,0} +(\cdot,{\tiny \yng(1)},\cdot)_{0,a}+(\cdot,\cdot,{\tiny \yng(1)})_{-(2k+a),-(N-2k-2a)} \\
{\tiny \overline{\yng(1)}} & \rightarrow ({\tiny \overline{\yng(1)}},\cdot,\cdot)_{-a,0} +(\cdot,{\tiny \overline{\yng(1)}},\cdot)_{0,-a}+(\cdot,\cdot,{\tiny \overline{\yng(1)}})_{2k+a,N-2k-2a},
\end{align}
where the Coulomb branch is associated with the $U(1)_1$ gauge symmetry whose CS level correctly vanishes. 
At the low-energy limit, the $SU(2k+a) \times SU(a)$ part becomes a pure CS gauge theory with no dynamical quark. The effective CS levels are shifted by integration of the massive components and are given by
\begin{align}
k_{eff}^{SU(2k+a)} =2k+a,~~~~~k_{eff}^{SU(a)} =-a,
\end{align}
which leads to a supersymmetric vacuum (non-zero Witten index) \cite{Intriligator:2013lca}. Therefore, this direction can be a quantum flat direction. However, due to the mixed CS term $k_{eff}^{U(1)_1, U(1)_2}$, the bare operator $Y^{bare}$ obtains a non-zero $U(1)_2$ charge. The gauge-invariant dressed monopole is defined by
\begin{align}
Y_d &:=Y^{bare}   \left[(\cdot,{\tiny \overline{\yng(1)}},\cdot)_{0,-a}  \right]^{N-2k-2a} \sim Y^{bare} \tilde{Q}^{N-2k-2a},
\end{align}
where the gauge indices of $\tilde{Q}^{N-2k-2a}$ are contracted by an epsilon tensor of the $SU(N-2k-2a)$ subgroup. As a result, the dressed operator becomes a flavor singlet. 
When $\bar{F}=N-2a -2k =0$, the mixed CS term $k_{eff}^{U(1)_1, U(1)_2}$ vanishes and the bare operator $Y^{bare} $ becomes gauge-invariant. We can also give another interpretation of the dressed monopole in a way that is consistent with the superconformal indices. 
In terms of the bare operator $Y_{SU(N-2)}^{bare}$, the dressed monopole is expressed as 
\begin{align}
Y_d &\sim Y_{SU(N-2)}^{bare} ({\tiny \overline{\yng(1)}}_{\,1,N})^{2k} ({\tiny \overline{\yng(1)}}_{\,-1,N}{\tiny \overline{\yng(1)}}_{\,1,N})^{a-1} ({\tiny \overline{\yng(1)}}_{\,0,2})^{N-2k-2a} \\
&\sim Y_{SU(N-2)}^{bare} W_\alpha^{2k} W_\alpha^{2(a-1)} \tilde{Q}^{N-2k-2a},
\end{align}
where the non-abelian color indices are totally anti-symmetrized by an epsilon tensor of the $SU(N-2)$ subgroup. Because of the power of $W_\alpha^{2(a-1)}$, there is no Coulomb branch for $a=0$. The absence of the monopole operators for $a=0$ will be more carefully studied in the next subsection.
The low-energy dynamics is described by a free meson and a single constraint $BY_d =1$. We can easily check that this constraint is consistent with the previous analysis via a complex mass deformation. This is also consistent with the duality which will be studied in the next section.

\begin{table}[H]\caption{The 3d $\mathcal{N}=2$ $SU(N)_{k}$ gauge theory with $(F,\bar{F})=(N,N-2a-2k)$} 
\begin{center}\scalebox{0.97}{
  \begin{tabular}{|c||c|c|c|c|c|c| } \hline
  &\footnotesize $SU(N)_{k}$&\footnotesize $SU(N)$&\footnotesize $SU(N-2a-2k)$&\footnotesize $U(1)$& \footnotesize $U(1)$& \footnotesize $U(1)_R$ \\ \hline
$Q$& ${\tiny \yng(1)}$&$\tiny \yng(1)$&1&$1$&0&$0$ \\  
$\tilde{Q}$& ${\tiny \overline{\yng(1)}}$&1&${\tiny \yng(1)}$&0&1&0  \\ \hline
$M:=Q\tilde{Q}$&1&${\tiny \yng(1)}$&${\tiny \yng(1)}$&$1$&1&$0$ \\ 
$B:=Q^N$&1&1&1&$N$&0&0  \\ \hline
$Y^{bare}$& \scriptsize $U(1)_2$: $a(N-2k-2a)$&1&1&$-N$& \scriptsize $-(N-2a-2k )$&0  \\
\scriptsize  $Y_d :=Y^{bare}\tilde{Q}^{N-2k-2a}$&1&1&1&$-N$&$0$&$0$  \\ \hline
  \end{tabular}}
  \end{center}\label{SU(N)QD}
\end{table}

\subsection{$SU(N)_k$ with $(F,\bar{F})=(N,N-2k)$}
Finally, we consider the 3d $\mathcal{N}=2$ $SU(N)_k$ gauge theory with $(F,\bar{F})=(N,N-2k)$ (anti-)fundamental quarks. This example is identical to the previous one (Table \ref{SU(N)QD}) but with $a=0$. As advocated above, there is no monopole operator for $a=0$. However, this absence of the Coulomb branch is more subtle than expected since there could be naively a lot of monopoles in the SCI expansion. We here explain why the SCI includes no contribution from the dressed monopole. As before, the bare Coulomb branch $Y^{bare}_{SU(N-2)}$ corresponds to the breaking $SU(N) \rightarrow SU(N-2) \times U(1)_1 \times U(1)_2$. Due to the (mixed) CS terms for the $U(1)_1$ and $U(1)_2$ gauge groups, the bare operator $Y^{bare}_{SU(N-2)}$ has the following $U(1)$ charges
\begin{align}
U(1)_1:~~-k,~~~~U(1)_2:~~~-k(N-2).
\end{align}
Notice that along the Coulomb branch, the matter contributions $\mathbf{1}_{ 1,N-2} \sim Q$ and ${\tiny \yng(1)}_{\, 0,-2}{\tiny \overline{\yng(1)}}_{\, 1,N} \sim Q W_\alpha $ have the correct $U(1)$ charges to dress the bare monopole. Therefore, we can define the following dressed operators
\begin{align}
&Y^{bare}_{SU(N-2)}  (\mathbf{1}_{ 1,N-2})^k \\
&Y^{bare}_{SU(N-2)}  (\mathbf{1}_{ 1,N-2})^{k-1}  ({\tiny \yng(1)}_{\, 0,-2}{\tiny \overline{\yng(1)}}_{\, 1,N}) \\
& \qquad \qquad \vdots \nonumber 
\end{align}
where $\mathbf{1}_{ 1,N-2}$ contributes as a fermion and ${\tiny \yng(1)}_{\, 0,-2}{\tiny \overline{\yng(1)}}_{\, 1,N} $ as a boson since the spins of the $U(1)_1$ charged components are transmuted under the monopole background \cite{Polyakov:1988md, Dimofte:2011py, Aharony:2015pla}. It is straightforward to see that these contributions beautifully cancel out each other. As a result, there is no Coulomb branch contribution in the SCI expansion. The low-energy dynamics is described by a free meson $M:=Q \tilde{Q}$ and a free baryon $B:=Q^N$. Although for other choices of $(N,F,\bar{F},k)$, we can construct various s-confinement phases with no monopole operator, we don't explicitly discuss them.

\section{$SU(N)_k$ Seiberg duality}
In this section, we investigate the SUSY Chern-Simons (so-called Giveon-Kutasov) duality for the $SU(N)$ gauge group. This was studied in \cite{Aharony:2014uya, Aharony:2013dha, Park:2013wta}. In \cite{Aharony:2014uya, Aharony:2013dha}, the baryonic branches and their matching under the duality were extensively studied. In this paper, we focus on the (dressed) monopole operators and then argue how they are transformed under the duality. For a special case, we can construct a duality between $SU(N)_k$ and $USp(2\tilde{N})$ gauge theories by employing the $USp$ CS duality \cite{Willett:2011gp, Benini:2011mf}. The form of the $SU(N)_k$ duality drastically changes depending on the sign of $k-\frac{1}{2}(F-\bar{F})$ \cite{Aharony:2014uya}. Through the studies of the dressed Coulomb branch, we will find that the duality should become complicated when $k=\frac{1}{2}(F-\bar{F})$ and that the duality reported in \cite{Aharony:2014uya} must be modified to have a correct matching of the baryon and Coulomb branch operators. 
  
\subsection{$SU(N)_{\frac{k}{2}}$ with $F-\bar{F} > k$}
Since the dual (magnetic) description drastically changes depending on the sign of $F-\bar{F} - k$ \cite{Benini:2011mf, Aharony:2014uya}, we start with the simplest case where the parameters of the theory satisfy 
\begin{align}
F-\bar{F} > k,~~~F>N,~~~\frac{1}{2}k+\frac{1}{2}(F-\bar{F}) \in  \mathbb{Z}.
\end{align}
The other parameter region will be studied in the next subsection. The electric description is given by a 3d $\mathcal{N}=2$ $SU(N)_{\frac{k}{2}}$ gauge theory with $F$ fundamental and $\bar{F}$ anti-fundamental chiral multiplets. The bare Chern-Simons level have to be half-odd integer when $F \pm \bar{F}$ is odd. The Higgs branch operators are defined by 
\begin{align}
M= Q \tilde{Q},~~~B:=Q^N,~~~\bar{B}:=\tilde{Q}^N, 
\end{align}
where the anti-baryon $\bar{B}$ is only available for $\bar{F} \ge N$. In what follows, we will focus on the Coulomb branch coordinates which are naively absent since the theory has a non-zero tree-level CS term.

As in the previous analysis, there could be two interpretations for the Coulomb branch coordinates. We first give the expression of the monopole operators which only include the low-energy massless modes. When the bare Coulomb branch (denoted by $Y^{bare}$) obtains a non-zero expectation value, the gauge group is spontaneously broken to
\begin{align}
SU(N) & \rightarrow SU(A) \times SU(N-A-B) \times SU(B) \times U(1)_1 \times U(1)_2   \label{SUCB}\\
{\tiny \yng(1)} & \rightarrow ({\tiny \yng(1)},\cdot,\cdot)_{B,0} +(\cdot,{\tiny \yng(1)},\cdot)_{0,B} +(\cdot, \cdot, {\tiny \yng(1)})_{-A,-(N-A-B)} \\
{\tiny \overline{\yng(1)}} & \rightarrow  ({\tiny \overline{\yng(1)}} ,\cdot,\cdot)_{-B,0} +(\cdot,{\tiny \overline{\yng(1)}} ,\cdot)_{0,-B} +(\cdot, \cdot, {\tiny \overline{\yng(1)}} )_{A,N-A-B},
\end{align}
where the Coulomb branch is constructed by dualizing the $U(1)_1$ vector superfield into a chiral superfield. The total number of the $U(1)$ factors above is reduced when some non-abelian factors vanish. The components charged under the $U(1)_1$ subgroup obtain real masses proportional to their $U(1)_1$ charges. Therefore, the low-energy theory has the following effective Chern-Simons terms 
\begin{align}
k_{eff}^{U(1)_1} &=AB \left[ \frac{1}{2}(A+B)k - \frac{1}{2} (A-B) (F-\bar{F}) \right] \\
k_{eff}^{U(1)_1,U(1)_2} &= AB(N-A-B) \left[ \frac{1}{2}k-\frac{1}{2} (F- \bar{F}) \right] \\
k_{eff}^{SU(A)} &= \frac{1}{2}k +\frac{1}{2} (F-\bar{F}) \\
k_{eff}^{SU(B)} &= \frac{1}{2}k -\frac{1}{2} (F-\bar{F}). 
\end{align}
Since we are studying the Coulomb branch associated with the $U(1)_1$ generator, the effective CS level $k_{eff}^{U(1)_1}$ (which is a mass term for the vector multiplet and lifts the moduli space) must be zero. For the $SU(A) \times SU(B)$ subgroup, its low-energy limit becomes a pure CS gauge theory without a dynamical quark. For this $SU(A) \times SU(B)$ CS theory to have a supersymmetric vacuum (and we also require $k_{eff}^{U(1)_1}=0$), we need to  take
\begin{align}
A= \frac{1}{2}k+\frac{1}{2}(F-\bar{F}),~~~~~B= -\frac{1}{2} k+\frac{1}{2}(F-\bar{F})
\end{align}
In the above calculation of the effective CS terms, we didn't list the CS terms like $k_{eff}^{SU(N-A-B)}$ since these additional CS terms don't give any constraint for possible $A$ and $B$.

In terms of the operators associated with the fundamental monopoles $Y_i$, the bare Coulomb branch operator is written as 
\begin{align}
Y^{bare} := \left( Y_1^B Y_2^{2B}  \cdots Y_A^{AB}Y_{A+1}^{AB} \cdots Y_{N-B}^{AB} Y_{N-B+1}^{A(B-1)} \cdots Y_{N-1}^A  \right)^{\frac{1}{AB}}.  \label{NAMSU}
\end{align}
As in the previous example, the monopole $Y^{bare}$ is a non-abelian monopole associated with the breaking \eqref{SUCB} and this leads to the fractional power in \eqref{NAMSU} \cite{Csaki:2014cwa, Amariti:2015kha, Nii:2018bgf}.
Due to the mixed CS term $k_{eff}^{U(1)_1,U(1)_2}$, the bare operator $Y^{bare}$ obtains a non-zero $U(1)_2$ charge. Therefore, we need to define a baryon-monopole operator as follows:
\begin{align}
Y_d := \left. Y^{bare}  \left[  (\cdot,{\tiny \overline{\yng(1)}} ,\cdot)_{0,-B} \right]^{N-A-B} \sim Y^{bare} \tilde{Q}^{N-A-B}\right|_{A= \frac{1}{2}k+\frac{1}{2}(F-\bar{F}),~B= - \frac{1}{2} k+\frac{1}{2}(F-\bar{F})},
\end{align}
where the color indices of $\tilde{Q}^{N-A-B}$ are contracted by an epsilon tensor of the $SU(N-A-B)$ subgroup. This operator is available only for $N \ge F-\bar{F}$.

This dressed operator can be interpreted in a different way that the viewpoint of the superconformal indices naturally does. This interpretation was pursued in \cite{Aharony:2015pla} for the $U(N)$ gauge group in detail. The bare monopole operator $ Y^{bare}_{SU(N-2)}$ is here defined by the gauge symmetry breaking
\begin{align}
SU(N) & \rightarrow SU(N-2) \times U(1)_1 \times U(1)_2 \\
{\tiny \yng(1)} & \rightarrow  {\tiny \yng(1)}_{ \,0,-2} +\mathbf{1}_{1,N-2} +\mathbf{1}_{-1,N-2} \\
{\tiny \overline{\yng(1)}} & \rightarrow {\tiny \overline{\yng(1)}}_{\, 0,2} +\mathbf{1}_{-1,-(N-2)} +\mathbf{1}_{1,-(N-2)}  \\
\mathbf{adj.} & \rightarrow \mathbf{adj.}_{0,0}+ \mathbf{1}_{0,0}+\mathbf{1}_{0,0} + {\tiny \yng(1)}_{ \,-1,-N}+{\tiny \yng(1)}_{ \,1,-N}+ {\tiny \overline{\yng(1)}}_{\, 1,N}+ {\tiny \overline{\yng(1)}}_{\, -1,N}+\mathbf{1}_{2,0}+\mathbf{1}_{-2,0},
\end{align}
where $\mathbf{adj.}$ is a gaugino field and the Coulomb branch corresponds to the $U(1)_1$ generator. 
The dressed monopole is defined for the Coulomb branch operator $Y_{SU(N-2)}^{bare}$ as
\begin{align}
Y_d &:= Y^{bare}_{SU(N-2)} ({\tiny \overline{\yng(1)}}_{ \, 0,2})^{\bar{F}-F+N} ({\tiny \overline{\yng(1)}}_{ \, 1,N})^k ({\tiny \overline{\yng(1)}}_{ \, 1,N}{\tiny \overline{\yng(1)}}_{ \, -1,N})^{\frac{F-\bar{F}-2-k}{2}}  \nonumber \\
&\sim Y^{bare}_{SU(N-2)} \tilde{Q}^{\bar{F}-F+N} W_\alpha^k W_{\alpha}^{F-\bar{F}-2-k}, \label{Ydressed}
\end{align}
where the color indices of $\tilde{Q}$'s and $W_\alpha$'s are contracted by an epsilon tensor of the unbroken $SU(N-2)$ group. Note that the gaugino contributions, ${\tiny \overline{\yng(1)}}_{ \, 1,N}$ and ${\tiny \overline{\yng(1)}}_{ \, -1,N}$, are transmuted to bosons on the monopole background by getting the anomalous spins \cite{Polyakov:1988md, Dimofte:2011py, Aharony:2015pla}.

\begin{table}[H]\caption{The 3d $\mathcal{N}=2$ $SU(N)_{\frac{k}{2}}$ gauge theory with $(F,\bar{F})$} 
\begin{center}
\scalebox{0.83}{
  \begin{tabular}{|c||c|c|c|c|c|c| } \hline
  &$SU(N)_{\frac{k}{2}}$&$SU(F)$&$SU(\bar{F})$&$U(1)$&$U(1)$&$U(1)_R$ \\[3pt] \hline
$Q$& ${\tiny \yng(1)}$&$\tiny \yng(1)$&1&$1$&0&$0$ \\  
$\tilde{Q}$& ${\tiny \overline{\yng(1)}}$&1&$\tiny \yng(1)$&0&1&0  \\ \hline
$M:=Q\tilde{Q}$&1&${\tiny \yng(1)}$&$\tiny \yng(1)$&$1$&1&$0$ \\ 
$B:=Q^N$&1& [$N$-th]&1&$N$&0&0  \\ 
$\bar{B}:=\tilde{Q}^N$ $(\bar{F} \ge N)$&1&1&[$N$-th]&$0$&$N$&0  \\  \hline
$Y^{bare}_{SU(N-2)}$&\begin{tabular}{l}
   \scriptsize $U(1)_1$: $-k$ \\
    \scriptsize $U(1)_2:$ $-(N-2)(F-\bar{F})$
  \end{tabular}&1&1&$-F$&$-\bar{F}$&  \scriptsize$F+\bar{F}-2N+2$ \\
  \scriptsize$Y_d:=Y^{bare}_{SU(N-2)} \tilde{Q}^{\bar{F}-F+N} W_\alpha^{F-\bar{F}-2}$&1&1&\footnotesize [$(\bar{F}-F+N)$-th]&$-F$&$N-F$&$2(F-N)$  \\ \hline
  \end{tabular}}
  \end{center}\label{SU(N)electricsimple}
\end{table}

For $F-\bar{F} >k$, the dual description can be easily obtained as follows (This derivation is different from the one adopted in \cite{Aharony:2014uya}.): We start from the $SU(N)_{0}$ SQCD theory with $F$ fundamental and $\bar{F}+k$ anti-fundamental quarks whose dual is given by an $SU(F- N)_0$ with $(F, \bar{F}+k)$ for $F-\bar{F} >k$, where both the electric and magnetic theories has no CS term. By introducing positive (negative) real masses for the $k$ anti-fundamental matters on the electric (magnetic) side, the electric description flows to the present theory in Table \ref{SU(N)electricsimple}. On the dual side, the real masses are mapped to the negative masses for the $k$ anti-fundamental (dual) quarks. Therefore, the low-energy magnetic description is given by a 3d $\mathcal{N}=2$ $SU(F-N)_{-\frac{k}{2}}$ gauge theory with a tree-level superpotential
\begin{align}
W_{mag} =M q \tilde{q}.
\end{align}
In this argument, $F-\bar{F}> k$ is important since for $F-\bar{F} < k$ where the dual gauge group becomes $SU(\bar{F}+k-N)$, the $k$ positive real masses on the electric sides are mapped to real masses for all the magnetic matters and we have to take a low-energy limit at a non-trivial point of the magnetic Coulomb branch. For $F-\bar{F} \le k$, the duality was studied in \cite{Aharony:2014uya} and we will discuss its Coulomb branch for $F-\bar{F} \le k$ in the next subsection.

Let us consider the magnetic Coulomb branch. We first give a definition of the Coulomb branch in a way that is consistent with the SCI computation. The bare Coulomb branch, denoted by $\tilde{Y}^{bare}_{SU(F-N-2)} $, induces the gauge symmetry breaking $SU(F-N) \rightarrow SU(F-N-2) \times U(1)_1 \times U(1)_2$. Along the Coulomb branch, the chiral multiplets become massive and generate the effective CS levels
\begin{align}
k_{eff}^{U(1)_1} =-k,~~~~k_{eff}^{U(1)_1,U(1)_2} =(F-N-2)(F -\bar{F})
\end{align}
Due to these CS terms, the bare operator $\tilde{Y}^{bare}_{SU(F-N-2)} $ obtains the non-zero $U(1)_1$ and $U(1)_2$ charges. Therefore, we need to define the dressed operator
\begin{align}
\bar{B} &\sim \tilde{Y}^{bare}_{SU(F-N-2)} ({\tiny \overline{\yng(1)}}_{ \, 0,2})^{\bar{F}-N} ({\tiny \overline{\yng(1)}}_{ \, -1,F-N})^k ({\tiny \overline{\yng(1)}}_{ \, 1,F-N}{\tiny \overline{\yng(1)}}_{ \, -1,F-N})^{\frac{F-\bar{F}-2-k}{2}}  \nonumber  \\
&\sim \tilde{Y}^{bare}_{SU(F-N-2)} \tilde{q}^{\bar{F}-N} w_\alpha^k w_{\alpha}^{F-\bar{F}-2-k},
\end{align}
where the color indices of the matter and gaugino fields are contracted by an $SU(F-N-2)$ epsilon tensor. From the quantum numbers of this dressed operator, we find that the dressed monopole on the magnetic side is identified with the anti-baryon $\bar{B}:=\tilde{Q}^N$. 
Notice that this dressing is only possible for $\bar{F} \ge N$ since the power of $\tilde{q}^{\bar{F}-N}$ becomes negative for $\bar{F} < N$, which is consistent with the electric picture.

Although this operator identification is valid, the effective CS term $k_{eff}^{U(1)_1}$ gives a topological mass term to the corresponding vector multiplet. Therefore, the flat direction $\tilde{Y}^{bare}_{SU(F-N-2)} $ is removed from the moduli space. In what follows, we will give an alternative description in terms of low-energy degrees of freedom. The bare Coulomb branch coordinate $\tilde{Y}^{bare}$, which becomes massless even after including quantum corrections, induces the following gauge symmetry breaking
\begin{align}
SU(F-N) & \rightarrow SU(A) \times SU(F-N-A-B) \times SU(B) \times U(1)_1 \times U(1)_2, 
\end{align}
where $A= -\frac{k}{2} +\frac{1}{2} (F - \bar{F})$ and $B= \frac{k}{2} +\frac{1}{2} (F - \bar{F})$. For these choices of $A$ and $B$, the effective CS term $k_{eff}^{U(1)_1}$ correctly vanishes as in the electric Coulomb branch. In terms of the fundamental monopole operators $\tilde{Y}_{i}~(i=1,\cdots,F-N-1),$ the bare operator $\tilde{Y}^{bare}$ is defined as
\begin{align}
\tilde{Y}^{bare} := \left. \left( \tilde{Y}_1^B  \tilde{Y}_2^{2B}  \cdots \tilde{Y}_A^{AB}   \tilde{Y}_{A+1}^{AB} \cdots  \tilde{Y}_{F-N-B}^{AB}  \tilde{Y}_{F-N-B+1}^{A(B-1)} \cdots  \tilde{Y}_{F-N-1}^A  \right)^{\frac{1}{AB}}    \right|_{ \substack{A=-\frac{k}{2} +\frac{1}{2} (F - \bar{F}) \\ B=\frac{k}{2} +\frac{1}{2} (F - \bar{F}) } },
\end{align}
where $\tilde{Y}^{bare}$ corresponds to a non-abelian monopole whose $U(1)$ generator is $Z_A \times Z_B \times U(1)_1 \subset SU(A) \times SU(B) \times U(1)_1$ \cite{Csaki:2014cwa, Amariti:2015kha, Nii:2018bgf}. 
Due to the mixed CS term between the $U(1)_1$ and $U(1)_2$ gauge groups, the bare operator $\tilde{Y}^{bare} $ obtains a non-zero $U(1)_2$ charge \cite{Intriligator:2013lca}. In order to make it gauge-invariant, we have to define the dressed operator  
\begin{align}
\bar{B}  & \sim \tilde{Y}^{bare}  \left( (\cdot, {\tiny \overline{\yng(1)}}, \cdot)_{0,-B}   \right)^{\bar{F} -N} \nonumber \\
&\sim \tilde{Y}^{bare}  \tilde{q}^{\bar{F}-N},
\end{align}
whose quantum numbers suggest that the dressed Coulomb branch on the magnetic side should be identified with the anti-baryon $\bar{B}:=\tilde{Q}^N$. Due to the power of $ \tilde{q}^{\bar{F}-N}$, this dressed operator is well-defined only for $\bar{F} \ge N $. This situation is consistent with the electric side.

\begin{table}[H]\caption{The magnetic $SU(F-N)_{-\frac{k}{2}}$ description dual to Table \ref{SU(N)electricsimple}} 
\begin{center}
\scalebox{0.79}{
  \begin{tabular}{|c||c|c|c|c|c|c| } \hline
  &$SU(F-N)_{-\frac{k}{2}}$&$SU(F)$&$SU(\bar{F})$&$U(1)$&$U(1)$&$U(1)_R$ \\[3pt] \hline
$q$& ${\tiny \yng(1)}$&$\tiny \overline{\yng(1)}$&1&$\frac{N}{F-N}$&0&$0$ \\  
$\tilde{q}$& ${\tiny \overline{\yng(1)}}$&1&$\tiny \overline{\yng(1)}$&$\frac{-F}{F-N}$&$-1$&2  \\
$M$&1&${\tiny \yng(1)}$&$\tiny \yng(1)$&$1$&1&$0$ \\  \hline
$B:=q^{F-N}$&1& [$N$-th]&1&$N$&0&0  \\ 
$Y_d \sim \tilde{q}^{F-N}$&1&1&\footnotesize [$(\bar{F}-F+N)$-th]&$-F$&$N-F$&$2(F-N)$  \\ \hline
$\tilde{Y}^{bare}_{SU(F-N-2)}$&\begin{tabular}{l}
   \scriptsize $U(1)_1$: $k$ \\
    \scriptsize $U(1)_2:$ $-(F-N-2)(F-\bar{F})$
  \end{tabular}&1&1&$\frac{F(\bar{F}-N)}{F-N}$&$\bar{F}$&  \scriptsize$-F-\bar{F}+2N+2$ \\
  \scriptsize$\bar{B} \sim \tilde{Y}^{bare}_{SU(F-N-2)} \tilde{Q}^{\bar{F}-N} \tilde{W}_\alpha^{F-\bar{F}-2}$&1&1&[$N$-th]&$0$&$N$&$0$  \\ \hline
  \end{tabular}}
  \end{center}\label{}
\end{table}

For $F=N+2$ (and then $\bar{F} < N+2-k$), the dual gauge group becomes $SU(2)$, which is identical to $USp(2)$. Therefore, we can further apply the $USp(2)$ CS duality known in \cite{Willett:2011gp, Benini:2011mf} and obtain the $USp(k+N+\bar{F}-2)_{\frac{k}{2}}$ second dual description. The theory has two gauge-singlets $B$ and $Y_d$ which are identified with the baryon and dressed monopole operators, respectively. The quantum numbers of the elementary fields are summarized in Table \ref{USpdual}. The theory has a tree-level superpotential
\begin{align}
W_{mag} = Bqq+ Y_d \tilde{q} \tilde{q},
\end{align}
which lifts a part of the mesonic operators on the magnetic side. 
The operator identification under the duality is manifest from Table \ref{USpdual} which reads $M \sim q \tilde{q}$. Since the $USp(2 \tilde{N})$ theory has a bare CS level, there is no Coulomb branch. This in turn confirms that we correctly capture the structure of the $SU(N)_k$ Coulomb branch.  

\begin{table}[H]\caption{The $USp(k+N+\bar{F}-2)_{\frac{k}{2}}$ second dual description for $F=N+2$} 
\begin{center}
\scalebox{0.95}{
  \begin{tabular}{|c||c|c|c|c|c|c| } \hline
  &$USp(k+N+\bar{F}-2)_{\frac{k}{2}}$&$SU(F=N+2)$&$SU(\bar{F})$&$U(1)$&$U(1)$&$U(1)_R$ \\[3pt] \hline
$q$& ${\tiny \yng(1)}$&$\tiny \yng(1)$&1&$-\frac{N}{2}$&0&$1$ \\  
$\tilde{q}$& ${\tiny \yng(1)}$&1&$\tiny \yng(1)$&$\frac{N+2}{2}$&$1$&$-1$  \\
$B$&1&  ${\tiny \overline{\yng(1,1)}}$&1&$N$&0&0  \\ 
$Y_d $&1&1&${\tiny \overline{\yng(1,1)}}$&$-(N+2)$&$-2$&$4$  \\[2pt] \hline
$M\sim q \tilde{q}$ &1&${\tiny \yng(1)}$&${\tiny \yng(1)}$&1&1&0  \\ \hline
  \end{tabular}}
  \end{center}\label{USpdual}
\end{table}

\subsection{$SU(N)_{\frac{k}{2}}$ with $F-\bar{F} < k$}
In this subsection, we consider the $SU(N)_{\frac{k}{2}}$ Giveon-Kutasov duality with $F-\bar{F} < k$ where the dual description becomes more complicated than the previous subsection \cite{Aharony:2014uya, Park:2013wta, Aharony:2013dha}. The electric description is completely the same as the previous one (see Table \ref{SUCSdualityele2}). The only difference is that there is no Coulomb branch operator for $F-\bar{F} \le k$. This is because it is impossible to simultaneously satisfy the three necessary conditions $k_{eff}^{U(1)_1}= 0$, $|k_{eff}^{SU(A)}| \ge A$ and $|k_{eff}^{SU(B)}|\ge B$ which correspond to the flatness of the $U(1)_1$ Coulomb branch and the existence of a supersymmetric vacuum for the low-energy $SU(A) \times SU(B)$ pure CS gauge theory. One might consider that the dressed Coulomb branch is constructible by using the bare monopole operator whose insertion is associated with the breaking $SU(N) \rightarrow SU(N-2) \times U(1)_1 \times U(1)_2$. However, this is not the case since the power in \eqref{Ydressed} becomes negative. These two approaches of constructing the Coulomb branch are completely consistent with each other also for $F-\bar{F} < k$. As a result, the moduli space is described by the Higgs branch coordinates $M:=Q \tilde{Q}$, $B:=Q^N$ and $\bar{B}:=Q^N$ listed in Table \ref{SUCSdualityele2}.

\begin{table}[H]\caption{The 3d $\mathcal{N}=2$ $SU(N)_{\frac{k}{2}}$ gauge theory with $(F,\bar{F})$ and $F-\bar{F} < k$} 
\begin{center}
\scalebox{1}{
  \begin{tabular}{|c||c|c|c|c|c|c| } \hline
  &$SU(N)_{\frac{k}{2}}$&$SU(F)$&$SU(\bar{F})$&$U(1)$&$U(1)$&$U(1)_R$ \\[3pt] \hline
$Q$& ${\tiny \yng(1)}$&$\tiny \yng(1)$&1&$1$&0&$0$ \\  
$\tilde{Q}$& ${\tiny \overline{\yng(1)}}$&1&$\tiny \yng(1)$&0&1&0  \\ \hline
$M:=Q\tilde{Q}$&1&${\tiny \yng(1)}$&$\tiny \yng(1)$&$1$&1&$0$ \\ 
$B:=Q^N$ $(F \ge N)$&1& [$N$-th]&1&$N$&0&0  \\ 
$\bar{B}:=\tilde{Q}^N$ $(\bar{F} \ge N)$&1&1&[$N$-th]&$0$&$N$&0  \\  \hline
  \end{tabular}}
  \end{center}\label{SUCSdualityele2}
\end{table}

The magnetic description is given by a 3d $\mathcal{N} =2$ $U ( \tilde{N} )_{-\frac{k}{2}, \frac{F+ \bar{F}}{2}-N}$ gauge theory with $F$ fundamental and $\bar{F}$ anti-fundamental chiral multiplets \cite{Aharony:2014uya, Park:2013wta, Aharony:2013dha}, where $\tilde{N}=\frac{F + \bar{F}}{2}+\frac{k}{2}-N$. Note that $\tilde{N}$ always becomes an integer. The theory also includes a gauge-singlet meson $M$ and a tree-level superpotential 
\begin{align}
W_{mag}= M q \tilde{q},
\end{align}
where $M$ is identified with the electric meson $M\sim Q \tilde{Q}$.
The CS term for the $SU(\tilde{N})$ gauge group is $-\frac{k}{2}$ while the CS term for the overall $U(1)$ factor is $\frac{F+ \bar{F}}{2}-N$. Since the gauge group is unitary (not special unitary), all the magnetic (anti-)baryon operators are not gauge-invariant. Hence, we have to be careful of matching the (anti-)baryons $B$ and $\bar{B}$ under the duality. As opposed to the electric side, there could be a Coulomb branch. As in the previous examples, we can have two interpretations where one approach is in harmony with the SCI viewpoint and the other one is consistent with the low-energy description.

For simplicity, we first construct dressed monopole operators by using massive components, which is an approach consistent with the SCI point of view. After doing this, we will define the same operators by using a low-energy (massless) language. When the bare operator, denoted by $X^{bare}_{U(1)\,\pm}$, obtains a non-zero expectation value, the gauge group is spontaneously broken as
\begin{align}
U(\tilde{N})& \rightarrow U(1)_{CB} \times U(\tilde{N}-1) \\
{\tiny \yng(1)}_{\, 1} &  \rightarrow \mathbf{1}_{1,0} +{\tiny \yng(1)}_{\, 0,1} \\
{\tiny \overline{\yng(1)}}_{ \, -1} & \rightarrow   \mathbf{1}_{-1,0}+{\tiny \overline{\yng(1)}}_{ \, 0,-1} \\
\mathbf{adj.}_0 & \rightarrow \mathbf{adj.}_{0,0}+ \mathbf{1}_{0,0}+{\tiny \yng(1)}_{\, -1,1}+ {\tiny \overline{\yng(1)}}_{ \, 1,-1}
\end{align}
where the $U(1)_{CB}$ subgroup is defined by a generator $t_{U(1)_{CB}}=\mathrm{diag.} (1,0,\cdots,0)$ and associated with the Coulomb branch flat direction. The subscript of $X^{bare}_{U(1)\,\pm}$ indicates the sign of the vev that induces the above breaking. By integrating out the massive components which are charged under $U(1)_{CB}$, the effective Chern-Simons terms are generated along the Coulomb branch as 
\begin{align}
k_{eff}^{U(1)_{CB}} = 1-\frac{k}{2} \pm \frac{1}{2} (F-\bar{F}),~~~~~k_{eff}^{U(1)_{CB},U(1) \subset U(\tilde{N}-1)}=1,
\end{align}
where we only listed the abelian CS factors for our purpose. 
Notice that the massive components don't shift the mixed CS term $k_{eff}^{U(1)_{CB},U(1) \subset U(\tilde{N}-1)}$ but the $U(1)_{CB}$ CS term $k_{eff}^{U(1)_{CB}}$.
In order to cancel these $U(1)$ (gauge) charges of $X^{bare}_{U(1)\,\pm}$, we need to define the dressed monopole operators
\begin{align}
X_{+}^d &:= X^{bare}_{U(1)\,+} ({\tiny \yng(1)}_{\, 0,1})^{F-N} ({\tiny \yng(1)}_{\, -1,1})^{\frac{k}{2}-1-\frac{F -\bar{F}}{2}} \nonumber \\
& \sim  X^{bare}_{U(1)\,+} q^{F-N}w_\alpha^{\frac{k}{2}-1-\frac{F - \bar{F}}{2}}  \label{Xplus} \\
X_{-}^d &:= X^{bare}_{U(1)\,-}({\tiny \overline{\yng(1)}}_{ \, 0,-1})^{\bar{F}-N} ({\tiny \overline{\yng(1)}}_{ \, 1,-1})^{\frac{k}{2}-1+\frac{F -\bar{F}}{2}}   \nonumber \\
& \sim X^{bare}_{U(1)\,-} \tilde{q}^{\bar{F}-N}  w_\alpha^{\frac{k}{2}-1+\frac{F - \bar{F}}{2}},
\end{align}
where the color indices are contracted by an epsilon tensor of the unbroken $U(\tilde{N}-1)$ gauge group. The quantum numbers of these composites are summarized in Table \ref{Flessk}. The baryon and anti-baryon operators on the electric side are identified as
\begin{align}
X_+^d  \sim B:= Q^N,~~~~~X_-^d  \sim \bar{B}:= \tilde{Q}^N.
\end{align}

The above construction of the dressed monopoles is consistent with the superconformal index which is a sum over all the states with possible GNO charges. The Coulomb branch $ X^{bare}_{U(1)\, \pm} $ comes from the states with the following GNO charges
\begin{align}
\ket{ \pm 1, \underbrace{0,\cdots,0}_{\tilde{N}-1} \,},
\end{align}
which are identical to the operator insertion of $ X^{bare}_{U(1)\, \pm} $. By dressing these bare states with matter fields, the physical gauge-invariant states are defined as
\begin{align}
 ({\tiny \yng(1)}_{\, 0,1})^{F-N} ({\tiny \yng(1)}_{\, -1,1})^{\frac{k}{2}-1-\frac{F -\bar{F}}{2}} \ket{ + 1, 0,\cdots,0 \,},~~~~~({\tiny \overline{\yng(1)}}_{ \, 0,-1})^{\bar{F}-N} ({\tiny \overline{\yng(1)}}_{ \, 1,-1})^{\frac{k}{2}-1+\frac{F -\bar{F}}{2}}  \ket{ - 1, 0,\cdots,0 \,}.
\end{align}
However, these expressions include massive components of the gaugino fields. In addition, the bare states themselves are massive due to the CS term $k_{eff}^{U(1)_{CB}}$. Therefore, these states should be reinterpreted in terms of purely low-energy (massless) degrees of freedom for describing the Coulomb moduli space.

In what follows, we reinterpret these operators as genuinely massless flat directions. For that purpose, we have to consider the Coulomb branch $X_{\pm ,U(P)}^{bare}$, which induces the gauge symmetry breaking
\begin{align}
U(C+P)_{k_{SU(C+P)},k_{U(1)}} & \rightarrow U(C)_{k_{SU(C+P)},k_{U(1)_C}} \times U(P)_{k_{SU(C+P)},k_{U(1)_P}} \\
{\tiny \yng(1)}_{\, 1} &  \rightarrow  ({\tiny \yng(1)}_{\,1}, \mathbf{1}_0)+(\mathbf{1}_0, {\tiny \yng(1)}_{\, 1}) \\
{\tiny \overline{\yng(1)}}_{ \, -1} & \rightarrow  ({\tiny \overline{\yng(1)}}_{ \, -1}, \mathbf{1}_0)+(\mathbf{1}_0 , {\tiny \overline{\yng(1)}}_{ \, -1}).
\end{align}
We will see that the Coulomb branch $X_{\pm,U(P)}^{bare}$ can survive all the quantum corrections by fine-tuning the value of $C$ \cite{Aharony:2013dha, Aharony:2014uya}. In this breaking, the UV CS terms are decomposed into 
\begin{gather*}
k_{U(1)_C} = \frac{Ck_{U(1)} + Pk_{SU(C+P)}}{C+P},~~~~k_{U(1)_P} = \frac{Pk_{U(1)} + Ck_{SU(C+P)}}{C+P}  \\
k^{mixed}_{U(1)_C,U(1)_P}  = \frac{1}{C+P} (k_{U(1)} -k_{SU(C+P)}).
\end{gather*}
This breaking is realized by the following vacuum expectation value of the adjoint scalar in the vector multiplet
\begin{align}
\braket{\phi_{adj.}} =\mathrm{diag.} (\underbrace{v,\cdots,v}_{C},0,\cdots,0)~~~~(v>0), \label{adjvevU}
\end{align}
where we first focus on the positive adjoint vevs labeled by $X_{+,U(P)}^{bare}$. For positive vevs $v>0$, the effective CS terms (including the shifts from the 1-loop diagrams of the massive fermions) are induced as
\begin{gather*}
k_{eff}^{U(1)_C} =-\frac{k}{2} +C +\frac{1}{2}(F-\bar{F})  ,~~~~k_{eff}^{U(1)_P} = -\frac{k}{2} +P   \\
k_{eff}^{SU(C)} = -\frac{k}{2} +\frac{1}{2} (F-\bar{F}),~~~~k_{eff}^{SU(P)} =-\frac{k}{2} \\
k_{eff}^{U(1)_C,U(1)_P} =1
\end{gather*}
Since the Coulomb branch $X_{\pm,U(P)}^{bare}$ corresponds to the $U(1)_C$ subgroup, we require $k_{eff}^{U(1)_C}=0$, which leads to $C=\frac{k}{2} -\frac{1}{2}(F-\bar{F})$ and $P=F-N$. 
Notice that the Coulomb branch with $v < 0$ leads to a non-vanishing $k_{eff}^{U(1)_C}$ for these choices of $C$ and $P$. In this way, the Coulomb branch $X_{+, U(F-N)}^{bare}$ which is associated with the gauge symmetry breaking $U(\tilde{N}) \rightarrow U \left(\frac{k}{2} -\frac{1}{2}(F-\bar{F}) \right) \times U(F-N)$, can become a massless flat direction. Due to the mixed CS term $k_{eff}^{U(1)_C,U(1)_P}$ between the $U(1)_C$ and $U(1)_P$ gauge symmetry, we have to define a dressed monopole operator
\begin{align}
X_{+}^{d} &:= X_{+,U(F-N)}^{bare} (\mathbf{1}_0, {\tiny \yng(1)}_{\, 1} )^{F-N} \nonumber \\
&\sim X_{+,U(F-N)}^{bare}  Q^{F-N},
\end{align}
where the color indices of $Q^{F-N}$ are contracted by an epsilon tensor of the $SU(P)$ gauge group. From the quantum numbers of this dressed operator, we can identify it with \eqref{Xplus}.

Next, we consider the Coulomb branch spanned by $C$ negative eigenvalues $v < 0$ in \eqref{adjvevU}. This Coulomb branch is denoted by $X_{-,U(C) \times U(P)}^{bare}$. The Coulomb branch here is associated with the (nagative) flat direction of $U(C)$. By solving $k_{eff}^{U(1)_C} \stackrel{!}{=} 0$, we find that the allowed values of $C$ and $P$ are given by
\begin{align}
C=\frac{k}{2} +\frac{1}{2}(F-\bar{F}),~~~~P=\bar{F}-N,
\end{align}
which is consistent with the analysis in \cite{Aharony:2013dha, Aharony:2014uya}.
In addition, the low-energy pure $SU(C)$ gauge theory safely has a supersymmetric vacuum because of the effective CS term 
\begin{align}
\left.  k_{eff}^{SU(C)} \right|_{C=\frac{k}{2} +\frac{1}{2}(F-\bar{F})} =-\frac{k}{2}-\frac{1}{2}(F-\bar{F})=-C.
\end{align}
Similarly, the mixed CS term becomes $k_{eff}^{U(1)_C,U(1)_P} =1$ and then the bare operator $X_{-,U(C) \times U(P)}^{bare}$ has a non-zero $U(1)_P$ charge. The gauge-invariant operator is given by
\begin{align}
X_{-}^d &:= \left. X_{-, U(C) \times U(P)}^{bare}  (\mathbf{1}_0, {\tiny \overline{\yng(1)}}_{\, -1} )^{\bar{F}-N} \right|_{C=\frac{k}{2} +\frac{1}{2}(F-\bar{F}) } \nonumber \\
& \left. \sim X_{-, U(C) \times U(P)}^{bare} \tilde{q}^{\bar{F}-N} \right|_{C=\frac{k}{2} +\frac{1}{2}(F-\bar{F})},
\end{align}
where the color indices of $\tilde{q}^{\bar{F}-N}$ are contracted by an epsilon tensor of the unbroken $U(P)$ gauge group. From the quantum numbers of these dressed operators, we find the operator matching
\begin{align}
B\sim X_{+}^{d},~~~~~\bar{B} \sim X_{-}^d.
\end{align}

\begin{table}[H]\caption{The magnetic $U \left(\frac{F + \bar{F}}{2}+\frac{k}{2}-N \right)_{-\frac{k}{2}, \frac{F+ \bar{F}}{2}-N}$ description dual to Table \ref{SUCSdualityele2}} 
\begin{center}
\scalebox{0.8}{
  \begin{tabular}{|c||c|c|c|c|c|c| } \hline
  &\small $U \left(\frac{F + \bar{F}}{2}+\frac{k}{2}-N \right)_{-\frac{k}{2}, \frac{F+ \bar{F}}{2}-N}$&$SU(F)$&$SU(\bar{F})$&$U(1)$&$U(1)$&$U(1)_R$ \\[3pt] \hline
$q$& ${\tiny \yng(1)}_{\,1}$&$\tiny \overline{\yng(1)}$&1&$\frac{2N-\bar{F}}{F+\bar{F}-2N}$&$\frac{-\bar{F}}{F+\bar{F}-2N}$&$1$ \\  
$\tilde{q}$& ${\tiny \overline{\yng(1)}}_{\, -1}$&1&$\tiny \overline{\yng(1)}$&$\frac{-F}{F+\bar{F}-2N}$&$\frac{2N-F}{F+\bar{F}-2N}$&1  \\
$M$&1&${\tiny \yng(1)}$&$\tiny \yng(1)$&$1$&1&$0$ \\  \hline
$X_+$&\begin{tabular}{l}
   \scriptsize $U(1)_{CB}$: $-(\frac{k}{2}-1-\frac{F-\bar{F}}{2})$ \\
    \scriptsize $U(1)_{NA}$: $+(\frac{F+\bar{F}}{2}+\frac{k}{2}-N-1)$
  \end{tabular}&1&1&$\frac{F(\bar{F}-N)}{F+\bar{F}-2N}$&$\frac{\bar{F}(F-N)}{F+\bar{F}-2N}$&  \scriptsize $-(\frac{F+\bar{F}}{2}+\frac{k}{2}-N-1)$ \\
  \footnotesize $X_+^d:=X_+ q^{F-N}w_\alpha^{\frac{k}{2}-1-\frac{F - \bar{F}}{2}}$&1&[$N$-th]&1&$N$&0&0  \\   \hline
  $X_-$&\begin{tabular}{l}
   \scriptsize $U(1)_{CB}$: $\frac{k}{2}-1-\frac{F-\bar{F}}{2}$ \\
    \scriptsize $U(1)_{NA}$: $-(\frac{F+\bar{F}}{2}+\frac{k}{2}-N-1)$
  \end{tabular}&1&1&$\frac{F(\bar{F}-N)}{F+\bar{F}-2N}$&$\frac{\bar{F}(F-N)}{F+\bar{F}-2N}$&  \scriptsize $-(\frac{F+\bar{F}}{2}+\frac{k}{2}-N-1)$ \\ 
 \footnotesize $X_-^d:=X_- \tilde{q}^{\bar{F}-N} w_\alpha^{\frac{k}{2}-1 +\frac{F-\bar{F}}{2}}$&1&1&[$N$-th]&0&$N$&0  \\  \hline
  \end{tabular}}
  \end{center}\label{Flessk}
\end{table}

\subsection{$SU(N)_{k}$ with $k=\frac{1}{2}(F-\bar{F})$}
So far, we have discussed the Seiberg-like duality in the 3d $\mathcal{N}=2$ $SU(N)_k$ Chern-Simons gauge theory with $k > \frac{1}{2}(F-\bar{F})$ or $k < \frac{1}{2}(F-\bar{F})$. We here argue that the duality with $k = \frac{1}{2}(F-\bar{F})$ is more complicated than the previous cases. In  the literature \cite{Aharony:2014uya}, the duality in this case is supposed to be the same as the duality with $k > \frac{1}{2}(F-\bar{F})$. However, this is incorrect: On the electric side of the previous duality, the baryon and anti-baryon operators (and also the corresponding moduli spaces) are available when $F \ge N$ and $\bar{F} \ge N$. However, on the magnetic side, the magnetic Coulomb branch which will be identified with these (anti-)baryons are only defined for $k - \frac{1}{2}(F-\bar{F}) >0$. When the equality is satisfied, one of the Coulomb branches discussed in the previous subsection is not available. We will here propose a correct duality for $k = \frac{1}{2}(F-\bar{F})$ and discuss how it is derived from known dualities.

The electric theory is a 3d $\mathcal{N}=2$ $SU(N)_{k=\frac{F-\bar{F}}{2}}$ gauge theory with $F$ fundamental and $\bar{F}$ anti-fundamental quarks. As in the previous subsection, the theory has no Coulomb branch. The low-energy physics is described by the Higgs branch operators $M:=Q \tilde{Q},~B:=Q^N$ and $\bar{B}:=\tilde{Q}^N$. Table \ref{SUCSdualityeleequality} summarizes the quantum numbers of the Higgs branch coordinates. For $\bar{F} <N$, the anti-baryon $\bar{B}$ is not defined.  

\begin{table}[H]\caption{The 3d $\mathcal{N}=2$ $SU(N)_{k=\frac{F-\bar{F}}{2}}$ gauge theory with $(F,\bar{F})$ (anti-)fundamental quarks} 
\begin{center}
\scalebox{1}{
  \begin{tabular}{|c||c|c|c|c|c|c| } \hline
  &$SU(N)_{k=\frac{F-\bar{F}}{2}}$&$SU(F)$&$SU(\bar{F})$&$U(1)_Q$&$U(1)_{\tilde{Q}}$&$U(1)_R$ \\[3pt] \hline
$Q$& ${\tiny \yng(1)}$&$\tiny \yng(1)$&1&$1$&0&$r$ \\  
$\tilde{Q}$& ${\tiny \overline{\yng(1)}}$&1&$\tiny \yng(1)$&0&1&$\bar{r}$  \\ \hline
$M:=Q\tilde{Q}$&1&${\tiny \yng(1)}$&$\tiny \yng(1)$&$1$&1&$r+\bar{r}$ \\ 
$B:=Q^N$ $(F \ge N)$&1& [$N$-th]&1&$N$&0&$Nr$  \\ 
$\bar{B}:=\tilde{Q}^N$ $(\bar{F} \ge N)$&1&1&[$N$-th]&$0$&$N$&$N\bar{r}$  \\  \hline
  \end{tabular}}
  \end{center}\label{SUCSdualityeleequality}
\end{table}

The magnetic description is given by a 3d $\mathcal{N}=2$ $U(F-N)_{-k,-k+\frac{1}{2}(F-N)}$ gauge theory with $F$ fundamental quarks $q$, $\bar{F}$ anti-fundamental quarks $\tilde{q}$, an electron $b$ and a meson singlet $M$. The tree-level Chern-Simons term for the non-abelian $SU(F-N)$ part is $k_{SU(F-N)}=-k$ whereas the abelian part is $k_{U(1)}=-k +\frac{1}{2}(F-N)$. The theory has a tree-level superpotential
\begin{align}
W_{mag}=Mq \tilde{q}+ \tilde{V}_{+},
\end{align}
where $\tilde{V}_+$ is a Coulomb branch coordinate corresponding to the gauge symmetry breaking
\begin{align}
U(F-N) & \rightarrow U(1) \times U(F-N-1).
\end{align}
This Coulomb branch is associated with the first $U(1)$ generator and the subscript of $\tilde{V}_+$ means that this breaking is generated by a positive vacuum expectation value. Along the breaking, the effective CS term for this Coulomb branch becomes
\begin{align}
k_{eff}^{U(1)} =-k+\frac{1}{2} +\frac{1}{2}(F-\bar{F}) -\frac{1}{2} =0,
\end{align}
which means that this flat direction acquires no topological mass from the CS term. Due to the tree-level superpotential introduced above, this flat direction is removed from the chiral ring. For a negative vev, we can similarly consider the Coulomb branch $\tilde{V}_-$ but the effective CS level cannot be zero in this case, which means that the flat direction labeled by $\tilde{V}_-$ becomes massive.

The quantum numbers of the magnetic fields are summarized in Table \ref{SUCSdualitymagequality}. However, the matching of these global $U(1)$ symmetries under the duality needs a careful treatment as follows: Since the magnetic gauge group has an abelian factor, there is a topological $U(1)_{top}$ symmetry. On the magnetic side, the $U(1)$ global symmetries must be linear combinations between the topological $U(1)_{top}$ and global $U(1)$ symmetries. The global $U(1)$ symmetries are acting on the matter fields as in Table \ref{SUCSdualitymagequality}. In this example, the electric global symmetries are mapped as 
\begin{align}
J_{U(1)_Q}^{ele} & := J_{U(1)_Q}^{mag, matter} \\
J_{ U(1)_{\tilde{Q}}  }^{ele} &:= J_{U(1)_{\tilde{Q}}}^{mag, matter} -\frac{ \bar{F} }{2} J^{mag}_{U(1)_{top}} \\
J_{U(1)_R}^{ele} &:=J_{U(1)_R}^{mag, matter} +\frac{F-N+1}{2} J^{mag}_{U(1)_{top}}
\end{align}
where the superscript ``$matter$'' means that the corresponding currents act only on the magnetic chiral superfields. $J^{mag}_{U(1)_{top}}$ is a topological $U(1)$ symmetry associated with the magnetic gauge group. Note that the operators constructed from the magnetic vector superfields are charged under $J^{mag}_{U(1)_{top}}$ while the magnetic matter fields are neutral under $J^{mag}_{U(1)_{top}}$. Since the electric $U(1)$ global symmetries are mapped by these transformation laws, the Coulomb branch associated with positive or negative vevs can play a different role.

Let us consider the magnetic Coulomb branch where the gauge group is spontaneously broken as  
\begin{align}
U(F-N)_{-k,-k+\frac{1}{2}(F-N)} & \rightarrow U(C) \times U(P),~~~C+P=F-N \\
k_{eff}^{U(1) \subset U(C)} &:= -k-\frac{1}{2} (F-\bar{F}) +C =-(F-\bar{F})+C  \\
k_{eff, mixed}^{U(1) \times U(1)} &:=\frac{1}{2} +\frac{1}{2}=1,
\end{align}
where the Coulomb branch is associated with the $U(1)$ gauge group in the $U(C)$ subgroup. We here assume that this breaking is induced by negative eigenvalues of the adjoint scalar. Therefore, we denote the bare Coulomb branch by $V_{-,bare}^{U(C) \times U(P)}$. Since the electron $b$ is massive along the Coulomb branch and is charged under both the $U(1) \subset U(C)$ and $U(1) \subset U(F-N-C)$ symmetries, the mixed CS term is also shifted as above. In this way, at the low-energy limit, the CS level is properly quantized as it should be. Since we are studying the flat directions coming from the magnetic vector multiplet, the effective $U(1)_C$ CS term, which behaves as a topological mass term for the vector multiplet, must be zero. Therefore, we obtain $C=F-\bar{F}$ and $P=\bar{F}-N$. Notice that the Coulomb branch $V_{+,bare}^{U(F-\bar{F}) \times U(\bar{F}-N)}$, which is associated with a positive vev, leads to a non-zero $k_{eff}^{U(1) \subset U(C)} $ and cannot be a flat direction.  
Due to the mixed CS terms $k_{eff, mixed}^{U(1) \times U(1)} $, the bare operator $V_{-,bare}^{U(F-\bar{F}) \times U(\bar{F}-N)}$ has a non-zero $U(1)_P$ charge \cite{Affleck:1982as, Intriligator:2013lca}. As a result, we need to define a dressed operator
\begin{align}
\bar{B} \sim V_{-,bare}^{U(F-\bar{F}) \times U(\bar{F}-N)} \tilde{q}^{\bar{F}-N},
\end{align}
which is identified with the anti-baryon $B:=\tilde{Q}^N$. This is consistent with all the symmetries listed in Table \ref{SUCSdualitymagequality}. 

\begin{table}[H]\caption{The magnetic $U(F-N)_{-k,-k+\frac{1}{2}(F-N)}$ gauge theory dual to Table \ref{SUCSdualityeleequality}} 
\begin{center}
\scalebox{0.87}{
  \begin{tabular}{|c||c|c|c|c|c|c| } \hline
  &\small $U(F-N)_{-k,-k+\frac{1}{2}(F-N)}$&\small  $SU(F)$& \small $SU(\bar{F})$&\small  $U(1)_Q$&\small  $U(1)_{\tilde{Q}}$&\small  $U(1)_R$ \\[3pt] \hline
$q$& ${\tiny \yng(1)}_{\, 1}$&$\tiny \yng(1)$&1&$-1$&$0$&$1-r$ \\  
$\tilde{q}$& ${\tiny \overline{\yng(1)}}_{\, -1}$&1&$\tiny \yng(1)$&0&$-1$&$1-\bar{r}$  \\ 
$b$&$\mathbf{1}_{-(F-N)}$&1&1&$F$&$0$&$N-F+Fr$  \\ 
$M$&1&${\tiny \yng(1)}$&$\tiny \yng(1)$&$1$&1&$r+\bar{r}$ \\  \hline
$B\sim q^{F-N} b$ &1& [$N$-th]&1&$N$&0&$Nr$ \\  \hline
$ \tilde{V}_{+}$&1&1&1&0&0&2  \\
 \scriptsize  $V_{-,bare}^{U(F-\bar{F}) \times U(\bar{F}-N)}$&$U(1)_P$: $\bar{F}-N$&1&1&$0$&$\bar{F}$&$N-\bar{F}+\bar{F}\bar{r}$  \\
 \scriptsize  $\bar{B} \sim V_{-,bare}^{U(F-\bar{F}) \times U(\bar{F}-N)} \tilde{q}^{\bar{F}-N}$ $(\bar{F} \ge N)$&1&1&[$N$-th]&$0$&$N$&$N\bar{r}$  \\  \hline
  \end{tabular}}
  \end{center}\label{SUCSdualitymagequality}
\end{table}

We finally discuss how the above duality appears from the known duality \cite{Aharony:2013dha}. In order to derive the duality for $k=\frac{1}{2}(F-\bar{F})$, we need to begin with the vector-like $SU(N)$ Seiberg duality with no CS term \cite{Aharony:2013dha} and introduce real masses to flow into the chiral theory discussed in this subsection. The electric description is a 3d $\mathcal{N}=2$ $SU(N)$ gauge theory with $F$ (anti-)fundamental flavors. The quantum numbers of the elementary fields and its duality are summarized in Table \ref{vectorlikeSQCDelemag}. On the electric side, we introduce positive real masses for $F_m$ anti-fundamental quarks. These real masses correspond to the background gauging of the $SU(F) \times U(1)$ symmetry acting only on anti-fundamental quarks. In terms of the diagonal generators of the flavor symmetry, the real masses are decomposed into 
\begin{align}
m_{\tilde{Q}} &=\mathrm{diag.} (\overbrace{0,\cdots,0}^{F-F_m},\overbrace{ m,\cdots,m}^{F_m}) \nonumber \\
&=\frac{m}{F} \left[ - \, \mathrm{diag.} (\overbrace{F_m,\cdots,F_m}^{F-F_m}, \overbrace{-(F-F_m),\cdots,-(F-F_m)}^{F_m}) +F_m \mathbf{1}  \right]
\end{align}
By integrating out the massive anti-quarks, we obtain the 3d $\mathcal{N}=2$ $SU(N)_{k=\frac{F_m}{2}}$ gauge theory with $F$ fundamental quarks and $\bar{F}=F-F_m$ anti-fundamental quarks, which satisfies $k=\frac{1}{2} (F-\bar{F})$.

The magnetic description is given by a 3d $\mathcal{N}=2$ $U(F-N)$ gauge theory with $F$ dual flavors and a pair of an electron and a positron. There are two gauge singlets $M$ and $Y$ which are identified with the electric meson $Q\tilde{Q}$ and Coulomb branch operators. The magnetic superpotential is 
\begin{align}
W_{mag} =Mq \tilde{q} +Yb \tilde{b} +\tilde{X}_{+} +\tilde{X}_-,
\end{align}
where $\tilde{X}_{\pm}$ are the magnetic Coulomb branch coordinates. See \cite{Aharony:2013dha} for the detail of this duality. On the magnetic side, the real masses are mapped as
\begin{align}
m_{q} &=-\frac{mF_m}{2F} \mathbf{1} \\
m_{\tilde{q}} &= \frac{m}{F}\left[\, \mathrm{diag.} (\overbrace{F_m,\cdots,F_m}^{F-F_m}, \overbrace{-(F-F_m),\cdots,-(F-F_m)}^{F_m}) -\frac{F_m}{2} \mathbf{1} \right] \\
m_{b}&=\frac{mF_m(F-N)}{2F},~~~~m_{\tilde{b}}=\frac{mF_m(F+N)}{2F},~~~~m_{Y}=-mF_m
\end{align}
For the meson singlet $M_i^j \sim Q_i\tilde{Q}^j$, the real masses are introduced only for $M_{i}^{j=F-F_m+1,\cdots,F}$. Since all the dynamical (dual) quarks obtain non-zero real masses, we have to take a low-energy limit at a non-trivial point of the magnetic Coulomb branch. By following the argument in \cite{Aharony:2013dha, Aharony:2014uya}, we find that the low-energy limit should be taken at $\tilde{\sigma}=\frac{mF_m}{2F} \mathbf{1}_{(F-N) \times (F-N)}$, where $\tilde{\sigma}$ is the adjoint scalar in the magnetic vector multiplet and corresponds to the overall $U(1)$ factor in the $U(F-N)$ gauge group. All the components of $q$, $\tilde{q}_{i=1,\cdots,F-F_m}$, the electron $b$ and the meson components $M_{i=1,\cdots,F}^{j=1,\cdots,F-F_m}$ remain massless along this Coulomb branch. The other fields and components become massive. The resulting theory is precisely the same as Table \ref{SUCSdualitymagequality}.

\begin{table}[H]\caption{The electric $SU(N)_{k=0}$ with $F(\, {\tiny \protect\yng(1)}+ \, {\tiny \overline{\protect\yng(1)}} \,)$ (top) and its dual (bottom)} 
\begin{center}
\scalebox{1}{
  \begin{tabular}{|c||c|c|c|c|c|c| } \hline
  &$SU(N)$&$SU(F)$&$SU(F)$&$U(1)_Q$&$U(1)_{\tilde{Q}}$&$U(1)_R$ \\ \hline
$Q$& ${\tiny \yng(1)}$&$\tiny \yng(1)$&1&$1$&0&$0$ \\  
$\tilde{Q}$& ${\tiny \overline{\yng(1)}}$&1&$\tiny \yng(1)$&0&1&$0$  \\ \hline
$M:=Q\tilde{Q}$&1&${\tiny \yng(1)}$&$\tiny \yng(1)$&$1$&1&$0$ \\ 
$B:=Q^N$ &1& [$N$-th]&1&$N$&0&$0$  \\ 
$\bar{B}:=\tilde{Q}^N$ &1&1&[$N$-th]&$0$&$N$&$0$  \\  \hline
$Y$&1&1&1&$-F$&$-F$&$2F-2N+2$  \\ \hline \hline
&$U(F-N)$&$SU(F)$&$SU(F)$&$U(1)_Q$&$U(1)_{\tilde{Q}}$&$U(1)_R$ \\ \hline
$q$&${\tiny \yng(1)}_{\,1}$&${\tiny \overline{\yng(1)}}$&1&$-\frac{1}{2}$&$-\frac{1}{2}$&1  \\
$\tilde{q}$&${\tiny \overline{\yng(1)}_{\, -1}}$&1&${\tiny \overline{\yng(1)}}$&$-\frac{1}{2}$&$-\frac{1}{2}$&1  \\
$b$&$\mathbf{1}_{-(F-N)}$&1&1&$\frac{F+N}{2}$&$\frac{F-N}{2}$&$N-F$  \\
$\tilde{b}$&$\mathbf{1}_{F-N}$&1&1&$\frac{F-N}{2}$&$\frac{F+N}{2}$&$N-F$  \\
$M$&1&${\tiny \yng(1)}$&$\tiny \yng(1)$&$1$&1&$0$ \\
$Y$&1&1&1&$-F$&$-F$&$2F-2N+2$  \\ \hline
$B:=bq^{F-N}$ &1& [$N$-th]&1&$N$&0&$0$  \\ 
$\bar{B}:=\tilde{b}\tilde{q}^{F-N}$ &1&1&[$N$-th]&$0$&$N$&$0$  \\  \hline
$\tilde{X}_{\pm}$&1&1&1&0&0&2  \\ \hline
  \end{tabular}}
  \end{center}\label{vectorlikeSQCDelemag}
\end{table}

\subsubsection{$SU(2)_{k=\frac{1}{2}}$ with $(F,\bar{F})=(3,2)$}
Let us examine several simple examples of the above duality by focusing on the cases with $N=2$. The first example is a 3d $\mathcal{N}=2$ $SU(2)_{k=\frac{1}{2}}$ gauge theory with three fundamental and two anti-fundamental quarks, which satisfies $k=\frac{1}{2}(F-\bar{F})$. Since the gauge group is $SU(2)$, there is no Coulomb branch as discussed in Section 2. This is because the tree-level CS term cannot be canceled by integrating out massive chiral multiplets. As a result, the theory only has the Higgs branch, which is described by 
\begin{align}
M=Q\tilde{Q},~~~~B:=Q^2,~~~~\bar{B}:=\tilde{Q}^2.
\end{align}
The quantum numbers of these fields are summarized in Table \ref{SU(2)_k12_32ele}.

\begin{table}[H]\caption{The 3d $\mathcal{N}=2$ $SU(2)_{k=\frac{1}{2}}$ gauge theory with $(F,\bar{F})=(3,2)$} 
\begin{center}
\scalebox{1}{
  \begin{tabular}{|c||c|c|c|c|c|c| } \hline
  &$SU(2)_{k=\frac{1}{2}}$&$SU(3)$&$SU(2)$&$U(1)_Q$&$U(1)_{\tilde{Q}}$&$U(1)_R$ \\[3pt] \hline
$Q$& ${\tiny \yng(1)}$&$\tiny \yng(1)$&1&$1$&0&$r$ \\  
$\tilde{Q}$& ${\tiny \overline{\yng(1)}}$&1&$\tiny \yng(1)$&0&1&$\bar{r}$  \\ \hline
$M:=Q\tilde{Q}$&1&${\tiny \yng(1)}$&$\tiny \yng(1)$&$1$&1&$r+\bar{r}$ \\ 
$B:=Q^N$ &1&  ${\tiny \overline{\yng(1)}}$ &1&$2$&0&$2r$  \\ 
$\bar{B}:=\tilde{Q}^N$ &1&1&1&$0$&$2$&$2\bar{r}$  \\  \hline
  \end{tabular}}
  \end{center}\label{SU(2)_k12_32ele}
\end{table}

The magnetic description\footnote{Since the electric gauge group is $SU(2)$, there is no difference between the fundamental and anti-fundamental representations. Therefore, we can regard the electric side as the $SU(2)_{k=\frac{1}{2}}$ gauge theory with $(F,\bar{F})=(5-a,a)$. By applying the chiral $SU(N)_k$ duality \cite{Aharony:2014uya} for $a \neq 2$, we can obtain other dual descriptions with different gauge groups. In this subsection, we only investigate the case where the theory satisfies $k=\frac{1}{2}(F-\bar{F})$.} is given by the 3d $\mathcal{N}=2$ $U(1)_{k=0}$ gauge theory with three (vector-like) flavors without a tree-level CS term. The theory has a tree-level superpotential
\begin{align}
W_{mag}=Mq \tilde{q} +\tilde{V}_+,
\end{align}
where $\tilde{V}_+$ is a Coulomb branch coordinate associated with a positive vev. Since the magnetic theory is vector-like, the two Coulomb branch coordinates, which are associated with positive and negative vevs, are quantum-mechanically flat. $\tilde{V}_+$ is removed by the superpotential and $\tilde{V}_-$ is identified with the anti-baryon. Notice that the asymmetric roles of the two monopole operators are possible because the global $U(1)$ symmetries on the magnetic side are mixed with the topological $U(1)_{top}$ symmetry.    

\begin{table}[H]\caption{The $U(1)_{k=0}$ magnetic theory dual to Table \ref{SU(2)_k12_32ele}} 
\begin{center}
\scalebox{1}{
  \begin{tabular}{|c||c|c|c|c|c|c| } \hline
  &$U(1)_{k=0}$&$SU(3)$&$SU(2)$&$U(1)_Q$&$U(1)_{\tilde{Q}}$&$U(1)_R$ \\[3pt] \hline
$q$& $+1$&${\tiny \overline{\yng(1)}}$&1&$-1$&0&$1-r$ \\  
$\tilde{q}$& $-1$&1&$\tiny \yng(1)$&0&$-1$&$1-\bar{r}$  \\ 
$b$&$-1$&1&1&3&0&$-1+3r$  \\
$M$&0&${\tiny \yng(1)}$&$\tiny \yng(1)$&$1$&1&$r+\bar{r}$ \\ \hline
$B\sim qb$ &0&  ${\tiny \overline{\yng(1)}}$ &1&$2$&0&$2r$  \\  \hline
$\tilde{V}_+$&0&1&1&0&0&2  \\
$\bar{B} \sim \tilde{V}_{-}$ &0&1&1&$0$&$2$&$2\bar{r}$  \\  \hline
  \end{tabular}}
  \end{center}\label{SU(2)_k12_32mag}
\end{table}

As a consistency check of our analysis, we will compute the superconformal indices from the electric and magnetic theories. We observed that both of the descriptions give the following result

\footnotesize
\begin{align}
I^{SU(2)_{\frac{1}{2}}}_{(F,\bar{F})=(3,2)} &=1+x \left(3 t^2+6 t u+u^2\right)+x^2 \left(6 t^4+\frac{-3 t-2 u}{t^3 u^2}+16 t^3 u+21 t^2 u^2+6 t u^3-\frac{6 t}{u}-\frac{6 u}{t}+u^4-13\right) \nonumber \\ & 
+x^3 \left(10 t^6+30 t^5 u+\frac{2 t+3 u}{t^4 u^3}+51 t^4 u^2+56 t^3 u^3-\frac{16 t^3}{u}+21 t^2 u^4-51 t^2+\frac{6}{t^2}+6 t u^5-\frac{6 u^3}{t} \right. \nonumber \\ &
 \left. -66 t u+\frac{6}{t u}+u^6-36 u^2+\frac{3}{u^2}\right)+x^4 \left(15 t^8+48 t^7 u+91 t^6 u^2+126 t^5 u^3-\frac{30 t^5}{u}+126 t^4 u^4 \right. \nonumber \\ &
 \left. -105 t^4+56 t^3 u^5-180 t^3 u+21 t^2 u^6-198 t^2 u^2+\frac{15 t^2}{u^2}+\frac{15 u^2}{t^2}+6 t u^7-\frac{6 u^5}{t}-120 t u^3+\frac{66 t}{u}  \right. \nonumber \\ &
 \qquad  \qquad  \left.+\frac{66 u}{t}+u^8-36 u^4+88\right)+\cdots,
\end{align}
\normalsize

\noindent where $t$ and $u$ are the fugacity parameters for the two global $U(1)$ symmetries counting the numbers of $Q$ and $\tilde{Q}$. The r-charges are set to be $r=\bar{r}=\frac{1}{2}$ for simplicity although the matching of the electric and magnetic indices can be checked for other choices of the r-symmetry. We here give the operator interpretation of the indices: The second term $x \left(3 t^2+6 t u+u^2\right)$ consists of three contributions $M$, $B$ and $\bar{B}$. At $O(x^2)$, the term $\frac{-3 t-2 u}{t^3 u^2} x^2$ corresponds to the dressed monopoles $Y_{U(1)} (Q+\tilde{Q}) $, where $Y_{U(1)}$ is a monopole operator associsted with the breaking $SU(2)_{k=\frac{1}{2}} \rightarrow U(1)_{k=1}$. Notice that the boson fields $Q+\tilde{Q}$ are transmuted into fermions under the monopole $Y_{U(1)}$ background \cite{Polyakov:1988md, Dimofte:2011py}.

\subsubsection{$SU(2)_{k=1}$ with $(F,\bar{F})=(3,1)$}
The next example is a 3d $\mathcal{N}=2$ $SU(2)_{k=1}$ gauge theory with three quarks and a single anti-quark, which also satisfies $k=\frac{1}{2} (F-\bar{F})$. The moduli space of the electric theory is described by $M:=Q\tilde{Q}$ and $B:=Q^2$. Notice that there is no anti-baryon and no Coulomb branch in this example. Table \ref{SU(2)_k1_31ele} summarizes the quantum numbers of these moduli fields.

\begin{table}[H]\caption{The 3d $\mathcal{N}=2$ $SU(2)_{k=1}$ gauge theory with $(F,\bar{F})=(3,1)$} 
\begin{center}
\scalebox{1}{
  \begin{tabular}{|c||c|c|c|c|c| } \hline
  &$SU(2)_{k=1}$&$SU(3)$&$U(1)_Q$&$U(1)_{\tilde{Q}}$&$U(1)_R$ \\[3pt] \hline
$Q$& ${\tiny \yng(1)}$&$\tiny \yng(1)$&$1$&0&$r$ \\  
$\tilde{Q}$& ${\tiny \overline{\yng(1)}}$&1&0&1&$\bar{r}$  \\ \hline
$M:=Q\tilde{Q}$&1&${\tiny \yng(1)}$&$1$&1&$r+\bar{r}$ \\ 
$B:=Q^2$ &1&  ${\tiny \overline{\yng(1)}}$ &$2$&0&$2r$  \\ \hline
  \end{tabular}}
  \end{center}\label{SU(2)_k1_31ele}
\end{table}

The magnetic description is given by the 3d $\mathcal{N}=2$ $U(1)_{-\frac{1}{2}}$ gauge theory with three positrons and two electrons. The theory has a tree-level superpotential
\begin{align}
W_{mag}=Mq \tilde{q} +\tilde{V}_+,
\end{align}
which distinguishes the two electrons into $\tilde{q}$ and $b$, reducing the global symmetry into the one listed in Table \ref{SU(2)_k1_31mag}. The mesonic branch is identified with $Q\tilde{Q} \sim M$ while the baryonic branch is mapped as $B:=Q^2 \sim qb$. In this example, the Coulomb branch $\tilde{V}_+$ has zero effective CS level but is removed by the superpotential. Table \ref{SU(2)_k1_31mag} summarizes the quantum numbers of the gauge-invariant operators. The other operator $\tilde{V}_-$ cannot be a part of the moduli space since the effective CS term is non-zero.

\begin{table}[H]\caption{The $U(1)_{k=-\frac{1}{2}}$ magnetic description dual to Table \ref{SU(2)_k1_31ele}} 
\begin{center}
\scalebox{1}{
  \begin{tabular}{|c||c|c|c|c|c| } \hline
  &$U(1)_{k=-\frac{1}{2}}$&$SU(3)$&$U(1)_Q$&$U(1)_{\tilde{Q}}$&$U(1)_R$ \\[3pt] \hline
$q$& $1$&${\tiny \overline{\yng(1)}}$&$-1$&0&$1-r$ \\  
$\tilde{q}$& $-1$&1&0&$-1$&$1-\bar{r}$  \\ 
$b$&$-1$&1&3&0&$-1+3r$\\
$M$&0&${\tiny \yng(1)}$&$1$&1&$r+\bar{r}$ \\ \hline
$B \sim qb$ &0&  ${\tiny \overline{\yng(1)}}$ &$2$&0&$2r$  \\  \hline
$\tilde{V}_+$&0&1&0&0&2  \\
$\tilde{V}_-$&$-1$&1&0&1&$\bar{r}$  \\ \hline
  \end{tabular}}
  \end{center}\label{SU(2)_k1_31mag}
\end{table}

Finally, we can test the matching of the superconformal indices under the duality. On both the electric and magnetic sides, we observed the same indices. The result is expanded as

\footnotesize
\begin{align}
I^{SU(2)_{1}}_{(F,\bar{F})=(3,1)} &=1+3 x \left(t^2+t u\right)+x^2 \left(6 t^4+8 t^3 u-\frac{1}{t^3 u}+6 t^2 u^2-\frac{3 t}{u}-\frac{3 u}{t}-10\right)+x^3 \left(10 t^6+15 t^5 u \right. \nonumber \\ &
 \left. +15 t^4 u^2+10 t^3 u^3-\frac{8 t^3}{u}+\frac{u}{t^3}-24 t^2+\frac{9}{t^2}-24 t u+\frac{9}{t u}-8 u^2+\frac{1}{u^2}\right)+x^4 \left(15 t^8+24 t^7 u  \right. \nonumber \\ &
 \left.+27 t^6 u^2+24 t^5 u^3-\frac{15 t^5}{u}+15 t^4 u^4-42 t^4-\frac{3}{t^4}-46 t^3 u-\frac{10}{t^3 u}-42 t^2 u^2+\frac{3 t^2}{u^2}+\frac{3 u^2}{t^2}-\frac{3}{t^2 u^2}  \right. \nonumber \\ &
 \left. -15 t u^3+\frac{24 t}{u}+\frac{24 u}{t}+34\right)+x^5 \left(21 t^{10}+35 t^9 u+42 t^8 u^2+42 t^7 u^3-\frac{24 t^7}{u}+35 t^6 u^4-64 t^6  \right. \nonumber \\ &
 \left. +21 t^5 u^5+\frac{3 (t+u)}{t^5 u^2}-72 t^5 u-72 t^4 u^2+\frac{6 t^4}{u^2}-64 t^3 u^3+\frac{36 t^3}{u}-\frac{8 u}{t^3}-24 t^2 u^4+36 t^2  \right. \nonumber \\ &
 \qquad  \qquad  \left. -\frac{24}{t^2}+\frac{6 u^3}{t}+36 t u-\frac{24}{t u}+36 u^2-\frac{8}{u^2}\right)+\cdots,
\end{align}
\normalsize

\noindent where the r-charges are set to be $r=\bar{r}=\frac{1}{2}$ for simplicity. The second term $3 x \left(t^2+t u\right)$ corresponds to the baryon and meson $B+M$. In the third term, $x^2 \left(-\frac{3 t}{u}-\frac{3 u}{t} -\frac{1}{t^3 u}\right)$ is interpreted as $Q \psi_{\tilde{Q}} + \psi_Q \tilde{Q} +Y_{U(1)}W_\alpha $, where $Y_{U(1)}$ induces the gauge symmetry breaking $SU(2)_{k=1} \rightarrow U(1)_{k=2}$ and the bare monopole is made gauge-invariant by dressing it with a gaugino. On the magnetic side, these can be regarded as $\tilde{V}_- q + (qb) \psi_M +\psi_b \tilde{q}$. Notice that on the monopole background $\tilde{V}_-$, the scalar $q$ is transmuted into a fermion and then $\tilde{V}_- q$ appears with a negative sign \cite{Dimofte:2011py}. This operator interpretation is consistent under the duality. 
For higher-order terms, we can give a similar operator interpretation or interpret them as products of these operators.

\subsubsection{$SU(2)_{k=\frac{3}{2}}$ with $(F,\bar{F})=(3,0)$}
The final example is a 3d $\mathcal{N}=2$ $SU(2)_{k=\frac{3}{2}}$ gauge theory with three fundamental quarks. This is a completely chiral theory and then there is no anti-baryon moduli space. On the electric side, the moduli space is described only by the baryon operator $B:=Q^2$. As in the previous examples, there is no Coulomb branch since the $SU(2)$ gauge group has a tree-level CS term $k_{SU(2)}=\frac{3}{2}$ which cannot be canceled along the Coulomb branch. Table \ref{SU(2)_k3o2_30ele} summarizes the quantum numbers of the elementary and moduli fields.

\begin{table}[H]\caption{The 3d $\mathcal{N}=2$ $SU(2)_{k=\frac{3}{2}}$ gauge theory with $(F,\bar{F})=(3,0)$} 
\begin{center}
\scalebox{1}{
  \begin{tabular}{|c||c|c|c|c| } \hline
  &$SU(2)_{k=\frac{3}{2}}$&$SU(3)$&$U(1)_Q$&$U(1)_R$ \\[3pt] \hline
$Q$& ${\tiny \yng(1)}$&$\tiny \yng(1)$&1&$r$ \\   \hline
$B:=Q^2$ &1&  ${\tiny \overline{\yng(1)}}$ &$2$&$2r$  \\ \hline
  \end{tabular}}
  \end{center}\label{SU(2)_k3o2_30ele}
\end{table}

The magnetic description becomes a 3d $\mathcal{N}=2$ $U(1)_{k=-1}$ gauge theory with three positrons $q$ and an electron $b$. The theory has a tree-level superpotential $W_{mag}=\tilde{V}_+$, where $\tilde{V}_+$ is one of the Coulomb branch operators associated with a positive eigenvalue. Notice that the $U(1)_R$ symmetry on the magnetic side is mixed with the topological $U(1)_{top}$ symmetry, which results in the asymmetry of the $U(1)_R$ charges for the two monopole operators $\tilde{V}_{\pm}$. Due to the effective CS terms, one of these operators, which is here $\tilde{V}_-$, obtains a non-zero $U(1)_{gauge}$ charge and must be dressed by matter multiplets. $\tilde{V}_+$ is gauge-invariant and could be a flat direction. However, it is lifted by the superpotential.  

\begin{table}[H]\caption{The magnetic $U(1)_{-1}$ gauge theory dual to Table \ref{SU(2)_k3o2_30ele}} 
\begin{center}
\scalebox{1}{
  \begin{tabular}{|c||c|c|c|c| } \hline
  &$U(1)_{k=-1}$&$SU(3)$&$U(1)_Q$&$U(1)_R$ \\[3pt] \hline
$q$& $+1$&$\tiny \overline{\yng(1)}$&$-1$&$1-r$ \\   
$b$&$-1$&1&3&$-1+3r$  \\ \hline
$B \sim qb$ &0&  ${\tiny \overline{\yng(1)}}$ &$2$&$2r$  \\ \hline
$\tilde{V}_+$&0&1&0&2  \\ 
$\tilde{V}_-$&$-2$&1&0&$0$  \\ \hline
  \end{tabular}}
  \end{center}\label{SU(2)_k3o2_30mag}
\end{table}

As a consistency check of the proposed duality, we compute the superconformal indices by using the electric and magnetic descriptions. We observed a nice agreement and the result becomes 

\footnotesize
\begin{align}
I^{SU(2)_{k=\frac{3}{2}}}_{(F,\bar{F})=(3,0)} &=1+3 t^2 x+\left(6 t^4-9\right) x^2+\left(10 t^6-15 t^2+\frac{9}{t^2}\right) x^3+\left(15 t^8-21 t^4-\frac{3}{t^4}+8\right) x^4  \nonumber \\ 
&+\left(21 t^{10}-27 t^6-12 t^2+\frac{6}{t^2}\right) x^5+\left(28 t^{12}-33 t^8-24 t^4-\frac{6}{t^4}+61\right) x^6  \nonumber \\ & +\left(36 t^{14}-39 t^{10}-36 t^6+\frac{1}{t^6}+57 t^2-\frac{51}{t^2}\right) x^7+\left(45 t^{16}-45 t^{12}-48 t^8+30 t^4+\frac{12}{t^4}+35\right) x^8+\cdots,
\end{align}
\normalsize

\noindent where $t$ is a fugacity parameter for the global $U(1)$ symmetry. The r-charge of $Q$ is set to be $r=\frac{1}{2}$ for simplicity. The second term $3 t^2 x$ represents the baryon operator $B:=Q^2 \sim qb$. Let us focus on $\frac{9x^3}{t^2}$ in the cubic order $O(x^3)$. On the electric side, this can be decomposed into $\frac{6x^3}{t^2}+\frac{3x^3}{t^2}$ and interpreted as $\psi_Q \psi_Q +Y_{U(1)} QW_\alpha$, where $\psi_Q$ is a fermion component of the electric quarks. $Y_{U(1)}$ is a monopole operator whose insertion leads to the gauge symmetry breaking $SU(2)_{k=\frac{3}{2}} \rightarrow U(1)_{k_{U(1)}=3}$. Since the bare monopole obtains a non-zero $U(1)$ charge, a gauge-invariant combination becomes $Y_{U(1)} QW_\alpha$. The coefficients denote the dimensions of the representations under the flavor $SU(3)$ symmetry. On the magnetic side, $\frac{6x^3}{t^2}+\frac{3x^3}{t^2}$ can be regarded as $\tilde{V}_- q^2+\psi_q \psi_b$. We can observe that the composite field constructed from the electric chiral multiplet is mapped to the dressed monopole and vice versa. The higher-order terms can be understood in a similar way. This confirms our analysis and the validity of the proposed duality.

\section{Summary}
In this paper, we studied the quantum structure of the Coulomb moduli space in the 3d $\mathcal{N}=2$ $SU(N)_k$ gauge theory with $F$ fundamental and $\bar{F}$ anti-fundamental matter fields. First, we studied the allowed monopole operators by focusing on the s-confinement phases, where we found that the breaking pattern of the gauge symmetry induced by the allowed bare monopole is quite different from the vector-like or chiral SQCD theory with no CS level. We also found that the (dressed) monopole operators which describe the Coulomb moduli space can be interpreted in two different ways: One approach is in harmony with the SCI computation in a sense that the dressed monopole is based on the GNO charges and written in terms of massive excitations. The other approach defines the dressed monopole solely in terms of massless excitations and is based on the non-abelian monopole whose magnetic charge is smaller than the minimal GNO charge. Therefore, the latter approach is consistent with the low-energy picture of the $SU(N)_k$ Chern-SImons matter theories.

We concretely exhibited the confinement phases for the $SU(2), SU(3), SU(4)$ and $SU(5)$ gauge groups. For the $SU(2)$ and $SU(3)$ cases, we computed the superconformal indices and observed a nice agreement between the electric and confining descriptions. In Section 6, we described the confinement phases for a generic $SU(N)_k$ gauge group by defining the (dressed) monopole operator.

In Section 7, we examined the Seiberg-like $SU(N)_k$ duality which is a generalization of the Giveon-Kutasov duality to special unitary gauge groups. In the literature, it was considered that there are two magnetic descriptions for $k < \frac{1}{2}(F-\bar{F})$ or $k \ge \frac{1}{2}(F-\bar{F})$ \cite{Aharony:2014uya}. However, we found that the third dual description must appear when $k= \frac{1}{2}(F-\bar{F})$ in order to have the correct matching between the electric (anti-)baryons and the magnetic monopole operators. We also derived the third magnetic description for $k= \frac{1}{2}(F-\bar{F})$ from the SQCD Seiberg duality \cite{Aharony:2013dha}. As a further support for this new duality, we computed the electric and magnetic superconformal indices for several $SU(2)_k$ cases and observed a beautiful agreement. In order to see the validity of this duality, it was important to mix the $U(1)_R$ symmetry with the topological $U(1)_{top}$ symmetry on the magnetic side. This mixing leads to the different roles of the two Coulomb branch operators $\tilde{X}_{\pm}$. Since the obtained duality is similar to the $U(N)$ CS duality with a monopole potential studied in \cite{Benini:2017dud}, it would be interesting to consider the connection between these dualities.   

\section*{Acknowledgements}
Keita Nii would like to thank Ofer Aharony, Francesco Benini and Shigeki Sugimoto for valuable comments. Keita Nii is the Yukawa Research Fellow supported by Yukawa Memorial Foundation.


\bibliographystyle{ieeetr}
\bibliography{3dCSM}

\end{document}